\documentclass[a4paper,11pt]{article}
\pdfoutput=1 % if your are submitting a pdflatex (i.e. if you have % images in pdf, png or jpg format)

\usepackage{jheppub} % for details on the use of the package, please
\usepackage[T1]{fontenc} % if needed

\usepackage{booktabs}    % some options for tabulars\\
\usepackage{placeins}
\usepackage{soul}
\usepackage[usenames, dvipsnames, svgnames]{xcolor}
\definecolor{redd}{rgb}{0.8, 0.1,0.2}
\definecolor{navy}{rgb}{0.05, 0.23,0.75}
\usepackage{bbm}
\usepackage{amsfonts}
\usepackage{amsmath,amssymb}
\usepackage{mathrsfs}
\usepackage{epsfig}
\usepackage{graphicx}
\usepackage{wrapfig}                 
\usepackage{url}
\usepackage{hyperref}
\usepackage{float}
\usepackage{color}
\usepackage{multirow}
\usepackage{lipsum}
\usepackage{enumitem}
\usepackage{subfigure}
\usepackage{tikz}
\usepackage{tikz-feynhand}
\usepackage[enableskew]{youngtab}
\usepackage{ytableau}
\usepackage{slashed}
\usepackage{ctable}
\newcolumntype{L}{>{\centering\arraybackslash}m{1.5cm}}

\usepackage{enumitem}

\usepackage{chngcntr}

\newcommand{\bear}{\begin{array}}
\newcommand {\eear}{\end{array}}

\newcommand{\beq}{\begin {equation}}
\newcommand{\eeq}{\end   {equation}}
\newcommand{\bea}{\begin {eqnarray}}
\newcommand{\eea}{\end   {eqnarray}}
\newcommand{\baa}{\begin {array}   }
\newcommand{\eaa}{\end   {array}   }
\newcommand{\bit}{\begin {itemize} }
\newcommand{\eit}{\end   {itemize} }
\newcommand{\be }{\begin {equation}}
\newcommand{\ee }{\end   {equation}}
\newcommand{\nn }{\nonumber        }

\def\bea{\begin{eqnarray}}
\def\eea{\end{eqnarray}}

\newcommand{\bef}{\begin{figure}}
\newcommand {\eef}{\end{figure}}
\newcommand{\bec}{\begin{center}}
\newcommand {\eec}{\end{center}}

\definecolor{cerulean}{rgb}{0., 0.62,0.7}

\newcommand{\twiddle}{{\raise.17ex\hbox{$\scriptstyle\sim$}}}

\title{Understanding the SM gauge group from SMEFT}
\preprint{IRMP-CP3-24-10}

\author[a]{Hao-Lin Li,}
\author[b]{Ling-Xiao Xu.}

\affiliation[a]{Centre for Cosmology, Particle Physics and Phenomenology (CP3), Universite Catholique de Louvain, Chem. du Cyclotron 2, 1348, Louvain-la-neuve, Belgium}
\affiliation[b]{Abdus Salam International Centre for Theoretical Physics, Strada Costiera 11, 34151, Trieste, Italy}

\emailAdd{haolin.li@uclouvain.be}
\emailAdd{lxu@ictp.it}

\abstract{
We discuss heavy particles that can be used to pin down the faithful Standard Model (SM) gauge group and their patterns in the SM effective field theory (SMEFT). These heavy particles are not invariant under a specific $\mathbb{Z}_6$ subgroup of $SU(3)_c\times SU(2)_L \times U(1)_Y$, which however acts trivially on all the SM particles, hence the faithful SM gauge group remains undetermined. Different realizations of the faithful SM gauge group correspond to different spectra of heavy particles, and they also correspond to distinct sets of line operators with one-form global symmetry acting on them. We show that the heavy particles not invariant under the $\mathbb{Z}_6$ group cannot appear in tree-level ultraviolet completions of SMEFT, this enforces us to consider one-loop UV completions of SMEFT to identify the $\mathbb{Z}_6$ non-invariant heavy particles. We demonstrate with examples that correlations between Wilson coefficients provide an efficient way to examine models with $\mathbb{Z}_6$ non-invariant heavy particles. Finally, we prove that all the scalars that can trigger electroweak symmetry breaking must be invariant under the $\mathbb{Z}_6$ group, hence they cannot be used to probe the faithful SM gauge group.
}

\begin{document}
\maketitle 
\flushbottom

%%%%%%%%%%%%%%%%%%%%%%%%%%%%%%%%%%%%%%%%%%
\section{Introduction}
\label{sec:intro}
%%%%%%%%%%%%%%%%%%%%%%%%%%%%%%%%%%%%%%%%%%

The Standard Model (SM) of particle physics is constructed based on the gauge group 
\be
\tilde{G}=SU(3)_c\times SU(2)_L \times U(1)_Y\ , 
\ee
with all the matter fields furnishing various chiral representations whose gauge anomalies are canceled in a non-trivial way. In this sense, the SM is extremely minimal and elegant. Most importantly, it is also phenomenologically successful. 
Nevertheless, there exists a specific $\mathbb{Z}_6$ subgroup of $\tilde{G}$ acting trivially on all the SM fields, as recently emphasized in~\cite{Tong:2017oea}.~\footnote{This issue was somehow rediscovered in~\cite{Tong:2017oea} in light of the connection between the global form of the gauge group and the spectrum of line operators, mainly inspired by~\cite{Aharony:2013hda}. However, the ambiguity of the SM gauge group has been known since a long time ago, see e.g. chapter 5.3 of~\cite{ORaifeartaigh:1986agb} and~\cite{Hucks:1990nw} for early references.}
As a result, the SM gauge group is not uniquely determined, i.e. the group $\tilde{G}$ is only the covering group, and the genuine SM gauge group which acts \emph{faithfully} is 
\bea
G=\frac{\tilde{G}}{\Gamma}=\frac{SU(3)_c\times SU(2)_L \times U(1)_Y}{\Gamma}\ , 
\label{eq:SM_gauge}
\eea
where $\Gamma$ is a subgroup of $\mathbb{Z}_6$ and it can be either $\mathbb{Z}_6, \mathbb{Z}_3, \mathbb{Z}_2$ or $1$.  Different choices of $\Gamma$ result in physically distinct theories~\cite{Aharony:2013hda}, our ultimate goal is to determine which $\Gamma$ is realized in nature. 
Notice that all the allowed matter fields must be in representations of $G$ rather than just the covering group $\tilde{G}$; in particular, they must be invariant under $\Gamma$~\cite{Aharony:2013hda}.  

When a specific $\Gamma$ is realized in nature, it implies non-trivial constraints on the allowed representations $(R_3, R_2, Q_Y)$ of $\tilde{G}$ in the entire theory. This applies not only to the fields within the SM, which is viewed as an effective theory in the infrared (IR), but also to all the beyond-the-SM (BSM) heavy resonances in the ultraviolet (UV). When these resonances have infinite masses, they are in one-to-one correspondence with the Wilson lines~\cite{Tong:2017oea, Aharony:2013hda}~\footnote{Particles correspond to local operators (i.e. they are points), naively. However, these local operators are not gauge invariant. This implies that particles give rise to a set of non-genuine local operators living at the end of the Wilson lines in the same representations of the gauge group. It means that particles can screen lines; conversely, the Wilson lines can be viewed as the worldlines of particles. Due to screening, the Wilson lines can break (and hence be equivalent to trivial lines) above the mass scale of particles. As a result, the line operators are only well-defined and nontrivial below the mass scale of the heavy particles. Accordingly, the one-form symmetry is viewed as a low-energy accidental/emergent symmetry. See e.g.~\cite{Schafer-Nameki:2023jdn, Bhardwaj:2023kri} for more detailed discussions.}, and the global symmetry acting on the lines is known as one-form global symmetry~\cite{Gaiotto:2014kfa}, which is a particular type of generalized symmetry. See e.g.~\cite{Cordova:2022ruw, Schafer-Nameki:2023jdn, Brennan:2023mmt, Bhardwaj:2023kri, Luo:2023ive, Shao:2023gho} for reviews on generalized symmetries from different perspectives.

Reversing the logic, if new particles are discovered in the future, it is possible to determine which $\Gamma$ is realized in nature with their transformation properties under the $\mathbb{Z}_6$ subgroup of $\tilde{G}$. See appendix.~\ref{app1:Z6} for more explanations on how the $\mathbb{Z}_6$ group (and its subgroups) acts on a field.  
There are four possible scenarios for all the UV resonances. 
\begin{enumerate}

      \item If all the heavy BSM fields are invariant under $\mathbb{Z}_6$, the group $\Gamma$ still remains undetermined as in the SM. However, if this turns out to be the case, it may be more proper to write the SM gauge group as $G=SU(3)_c\times SU(2)_L \times U(1)_Y/\mathbb{Z}_6$.

      \item If there exists at least one heavy field in a representation which is not invariant under $\mathbb{Z}_3$ but invariant under $\mathbb{Z}_2$ (hence not invariant under $\mathbb{Z}_6$), $\Gamma$ can be either $\mathbb{Z}_2$ or $1$. 
      
      For example, one may consider $(R_3, R_2, Q_Y)=(\tiny\yng(1),\ \tiny\yng(1)\ , 1/2)$, which is allowed when $\Gamma=1$ or $\mathbb{Z}_2$, but forbidden when $\Gamma=\mathbb{Z}_3$ or $\mathbb{Z}_6$.

      \item If there exists at least one heavy field in a representation which is not invariant under $\mathbb{Z}_2$ but invariant under $\mathbb{Z}_3$ (hence not invariant under $\mathbb{Z}_6$), $\Gamma$ can be either $\mathbb{Z}_3$ or $1$. 
      
      For example, $(R_3, R_2, Q_Y)=(\tiny\yng(1),\ \tiny\yng(1)\ , 2/3)$ is only allowed when $\Gamma=1$ or $\mathbb{Z}_3$, but forbidden when $\Gamma=\mathbb{Z}_2$ or $\mathbb{Z}_6$.

      \item If there exists at least one heavy field in a representation invariant under neither $\mathbb{Z}_3$ nor $\mathbb{Z}_2$ (hence not invariant under $\mathbb{Z}_6$), $\Gamma$ is uniquely determined to be $1$. 
      
      For example, a heavy resonance in the representation $(R_3, R_2, Q_Y)=(\tiny\yng(1),\ \tiny\yng(1)\ , 0)$ is allowed when $\Gamma=1$, but forbidden when $\Gamma=\mathbb{Z}_6, \mathbb{Z}_3$, or $\mathbb{Z}_2$.

\end{enumerate}
In this paper, we call these heavy particles the ``\emph{$\mathbb{Z}_6$ exotics}'' if they transform nontrivially under the $\mathbb{Z}_6$ group.
In more realistic setups, some well-known examples of $\mathbb{Z}_6$ exotics include the original KSVZ fermions in QCD axion models~\cite{Kim:1979if, Shifman:1979if}, pure milli-charged particles and fractionally-charged particles~\cite{Perl:2009zz, Langacker:2011db}.
See also~\cite{DiLuzio:2016sbl, DiLuzio:2017pfr} for a comprehensive analysis on KSVZ fermions in axion models,~\cite{CMS:2012xi, CMS:2013czn, Dolgov:2013una, CDMS:2014ane, Haas:2014dda, Vinyoles:2015khy, Majorana:2018gib, Afek:2020lek, CMS-PAS-EXO-19-006, Foroughi-Abari:2020qar, Gan:2023jbs, ATLAS:2023zxo} for recent experimental and phenomenological studies on fractionally-charged and milli-charged particles, and~\cite{Cacciapaglia:2015yra, Murgui:2021eqf, Chung:2023iwj} for concrete BSM models involving $\mathbb{Z}_6$ exotic particles. Fractionally charged particles are also well motivated in superstring models~\cite{Wen:1985qj, Athanasiu:1988uj}. 
One striking phenomenological feature is that the lightest $\mathbb{Z}_6$ exotic must be cosmologically stable, and they are powerful probes for the reheating temperature~\cite{Gan:2023jbs}.

It turns out that discovering BSM heavy resonances is essential to understanding better the SM itself, namely the genuine SM gauge group. Conversely, it is also well-motivated to use the precise form of the SM gauge group as a guiding principle to classify the vast landscape of BSM physics. 
We notice that there have been similar theoretical and phenomenological interests in recent literature on the precise form of the SM gauge group; see~\cite{Davighi:2019rcd, Wan:2019gqr, Davighi:2020bvi} for discussions on non-perturbative gauge anomalies,~\cite{Anber:2021upc} on fractional instanton effects, and~\cite{Choi:2023pdp, Reece:2023iqn, Cordova:2023her} on axion coupling quantization. From a bottom-up perspective, we do not assume any UV inputs (such as the requirement of grand unification~\cite{Georgi:1974sy, Pati:1974yy}), which can otherwise uniquely determine $\Gamma$ and accordingly restrict the allowed representations. Instead, we explore phenomenological imprints at lower energies and allow experimental discoveries to guide us. If $\mathbb{Z}_6$ exotics are discovered, some scenarios of grand unification can be excluded. For example, grand unification based on $SU(5)$ group predicts $\Gamma=\mathbb{Z}_6$, which is falsified if any $\mathbb{Z}_6$ exotic particle is discovered.

Another important motivation to study the precise form of the SM gauge group is that theories with different realizations of $\Gamma$ admit different spectra of magnetically-charged heavy particles~\cite{Aharony:2013hda, Tong:2017oea}. They are in one-to-one correspondence with 't Hooft lines if their masses are infinite. It is intriguing to study their interactions with the SM particles systematically, particularly in light of the recent results on the fermion-monopole scatterings~\cite{Brennan:2021ewu, Kitano:2021pwt, Hamada:2022eiv,  Csaki:2022tvb,  Csaki:2022qtz, Brennan:2023tae, Khoze:2023kiu, vanBeest:2023dbu, vanBeest:2023mbs}, and possibly with dark sectors being magnetically charged under the SM~\cite{Terning:2018lsv, Terning:2019bhg, Graesser:2021vkr, Hiramatsu:2021kvu, Chitose:2023bnd}.  

When there is a large separation between the new physics scale and the weak scale, the Standard Model effective field theory (SMEFT) provides a powerful and systematic framework to parameterize the UV physics at low energy.  See~\cite{Isidori:2023pyp} for a recent review on SMEFT.  
%we aim to find more applications of generalized global symmetries~\cite{Gaiotto:2014kfa} in particle phenomenology. 
In this paper, we use the SMEFT to systematically characterize the heavy particles that are the smoking gun signatures for distinguishing different $\Gamma$. 
We demonstrate that $\mathbb{Z}_6$ exotics cannot appear in any tree-level UV completion of SMEFT. This result is valid for effective operators at all mass dimensions, forcing us to consider loop-level UV completions.  
This provides a strong motivation for the development of the computational tools for the one-loop matching automation~\cite{DasBakshi:2018vni, Carmona:2021xtq, Fuentes-Martin:2022jrf, Aebischer:2023nnv}. 
As an illustration, we consider two benchmark models containing scalar and fermionic $\mathbb{Z}_6$ exotics respectively, and perform the one-loop matching to obtain the Wilson coefficients in terms of parameters in the UV Lagrangian. 
We find that for these two classes of models, one can solve for the quantum numbers of the corresponding $\mathbb{Z}_6$ exotics in terms of the Wilson coefficients obtained from one-loop matching. This provides a novel strategy to examine UV models due to the discreteness of the $SU(3)_c$ and $SU(2)_L$ quantum numbers.

The paper is organized as follows. 
In section~\ref{sec:toymodel}, we analyze a toy model to illustrate how to derive the constraints on the allowed representations given the gauge group. In section~\ref{sec:SM_gauge_group}, we discuss the ambiguity in defining the genuine SM gauge group, the heavy resonances useful for removing this ambiguity, and their patterns in SMEFT. Here we emphasize the importance of one-loop matching in SMEFT. Some simple examples are also presented. 
In section~\ref{sec:pheno}, we show the general strategy to probe the models of $\mathbb{Z}_6$ exotic particles using Wilson coefficients in SMEFT.
In section~\ref{sec:non-decoupling_case}, we extend the analysis to non-decoupling scalars in a general electroweak symmetry breaking (EWSB) sector. We prove that all these scalars are not $\mathbb{Z}_6$ exotics. 
Finally, we conclude in section~\ref{sec:conclusion}. Some technical details are collected in appendices~\ref{app1:Z6},~\ref{app:derivation}. 

%%%%%%%%%%%%%%%%%%%%%%%%%%%%%%%%%%%%%%%%%%%%%%%%%%%%%%%%%%%%%%%%%%%%%%%%
\section{A toy model: $SU(2)$ versus $SO(3)$ gauge theories}
\label{sec:toymodel}
%%%%%%%%%%%%%%%%%%%%%%%%%%%%%%%%%%%%%%%%%%%%%%%%%%%%%%%%%%%%%%%%%%%%%%%%

Before we analyze the SM, it is useful to illustrate the methodology in a simpler toy model, based on the comparison between $SU(2)$ and $SO(3)$ gauge theories. 

Even though $SU(2)$ and $SO(3)$ have the same Lie algebra, they are distinct Lie groups differing by the global structure of the group manifold, i.e. $SO(3)$ is isomorphic to $SU(2)/\mathbb{Z}_2$ where $\mathbb{Z}_2$ is the center of $SU(2)$. The $\mathbb{Z}_2$ quotient implies that one can obtain all the allowed representations in the $SO(3)$ theory by starting with the representations of $SU(2)$ and removing the ones not invariant under the $\mathbb{Z}_2$ center of $SU(2)$.
As a result, $SU(2)$ and $SO(3)$ gauge theories are physically distinct theories differing by the allowed spectrum of matter fields (i.e. line operators if the particles are infinitely heavy)~\cite{Aharony:2013hda}.  
To be specific, $SU(2)$ gauge theory allows for matter fields in both half-integer and integer spin representations (i.e., Young diagram of $SU(2)$ with odd and even number of boxes), while $SO(3)$ gauge theory only allows for matter fields in integer spin representations (i.e., Young diagram of $SU(2)$ with only even number of boxes). The half-integer spin representations are excluded in the $SO(3)$ gauge theory because of the  $\mathbb{Z}_2$ quotient.

Suppose at low energy all the discovered matter fields are in integer spin representations of $\tilde{G}=SU(2)$, one cannot conclude the genuine gauge group is $SU(2)$. Instead, there are two options: the genuine gauge group is either $SU(2)$ or $SO(3)$. In other words, the genuine gauge group can be written as
\be
G=\frac{SU(2)}{\Gamma}\ ,
\ee
where $\Gamma$ is either $1$ or $\mathbb{Z}_2$. If any ``$\mathbb{Z}_2$ exotic" heavy particle, i.e. a particle in half-integer spin representation of $SU(2)$, is discovered, the gauge group is
determined to be $SU(2)$.

Suppose such a $\mathbb{Z}_2$ exotic particle is heavy and has a decoupling limit. In that case, it is natural to integrate it out and use effective theories to explore the low-energy phenomenological consequences. It is easy to observe that this $\mathbb{Z}_2$ exotic particle cannot appear in any tree-level UV completions of high dimensional operators. Hence it is necessary to perform matching at the loop level to integrate out the $\mathbb{Z}_2$ exotic particle in a UV theory. 

In the following, we apply the same methodology in the SM, with a more detailed analysis of heavy resonances in the Warsaw basis~\cite{Grzadkowski:2010es} of SMEFT, if they can decouple from the weak scale. As we will see, the SM reflects the same features as we illustrated above in the toy model.

%%%%%%%%%%%%%%%%%%%%%%%%%%%%%%%%%%%%%%%%%%%%%%%%%%%%%%%%%%%
\section{The SM gauge group, heavy resonances, and SMEFT}
\label{sec:SM_gauge_group}
%%%%%%%%%%%%%%%%%%%%%%%%%%%%%%%%%%%%%%%%%%%%%%%%%%%%%%%%%%%

%%%%%%%%%%%%%%%%%%%%%%%%%%%%%%%%%%%%%%%%
\subsection{Which Standard Model?}
%%%%%%%%%%%%%%%%%%%%%%%%%%%%%%%%%%%%%%%%%

It was recently emphasized in~\cite{Tong:2017oea} that there is a particular $\mathbb{Z}_6$ subgroup of $\tilde{G}=SU(3)_c\times SU(2)_L \times U(1)_Y$, generated by the centers of $SU(3)_c$ and $SU(2)_L$ combined with a specific rotation of $U(1)_Y$, acting trivially on all the SM fields. (See also chapter 5.3 of~\cite{ORaifeartaigh:1986agb} and~\cite{Hucks:1990nw} for early references which made the same point.)
For completeness, we also provide a lightening review in appendix~\ref{app1:Z6} to elucidate the $\mathbb{Z}_6$ group and how it acts on a general representation $(R_3, R_2, Q_Y)$ of $\tilde{G}=SU(3)_c\times SU(2)_L \times U(1)_Y$. Similar discussions can also be found in~\cite{Davighi:2019rcd, Wan:2019gqr, Choi:2023pdp, Reece:2023iqn, Cordova:2023her}. 

As we mentioned in section~\ref{sec:intro}, there are four possible SM gauge groups, i.e. eq.~\eqref{eq:SM_gauge}. But nature can only realize one of them. In particular, different realizations of $\Gamma$ result in physically distinct theories of the Standard Models, since one-form global symmetries distinguish them. From a particle physics perspective, this implies that the allowed representations for heavy particles are different for different $\Gamma$. 

In a snapshot, the analysis in appendix~\ref{app1:Z6} suggests the following four cases, which can be matched to the discussions in the introduction. In the following equations, $\mathcal{N}(R_3)$ and $\mathcal{N}(R_2)$ are respectively the N-alities~\footnote{N-ality of a representation of $SU(N)$ group equals the number of boxes of the corresponding Young diagram modulo $N$.} for the representations $R_3$ and $R_2$ under $SU(3)_c$ and $SU(2)_L$, and $6 Q_Y$ is the generator of $U(1)_Y$, with the hypercharge of the left-hand quark doublet $Q_L$ being normalized to $1/6$.
\begin{enumerate}
\item When $\Gamma=1$, all the representations $\mathcal{R}=(R_3, R_2, Q_Y)$ of $\tilde{G}$ are allowed. As a trivial group, the identity always acts trivially on any $(R_3, R_2, Q_Y)$, hence $\Gamma=1$ can never be excluded. 
\item When $\Gamma=\mathbb{Z}_2$, all the allowed representations $\mathcal{R}=(R_3, R_2, Q_Y)$ have to satisfy the constraint 
\bea
\label{eq:quantization_Z6_main_2}
\mathcal{N}(R_2) &=& 6 Q_Y\ \text{mod}\ 2\ , 
\eea
while $R_3$ can be any representation of $SU(3)_c$. This is the condition when $\mathbb{Z}_2$ acts trivially on $\mathcal{R}$. Notice that the SM gauge group can also be written as $SU(3)_c\times U(2)_L$ when $\Gamma=\mathbb{Z}_2$, where $U(2)_L=SU(2)_L\times U(1)_Y/\mathbb{Z}_2$. 
\item When $\Gamma=\mathbb{Z}_3$, all the allowed representations $\mathcal{R}=(R_3, R_2, Q_Y)$ have to satisfy the constraint 
\bea
\label{eq:quantization_Z6_main_1}
\mathcal{N}(R_3) &=& 6 Q_Y\ \text{mod}\ 3\ , 
\eea
while $R_2$ can be any representation of $SU(2)_L$. This is the condition when $\mathbb{Z}_3$ acts trivially on $\mathcal{R}$. Similar to the previous case, the SM gauge group can also be written as $U(3)_c\times SU(2)_L$ when $\Gamma=\mathbb{Z}_3$, where $U(3)_c=SU(3)_c\times U(1)_Y/\mathbb{Z}_3$.
\item When $\Gamma=\mathbb{Z}_6$, both of its nontrivial subgroups $\mathbb{Z}_2$ and $\mathbb{Z}_3$ have to act trivially on $\mathcal{R}=(R_3, R_2, Q_Y)$, hence both constraints of eqs.~\eqref{eq:quantization_Z6_main_2} and~\eqref{eq:quantization_Z6_main_1} need to be satisfied. 

One may also write the SM gauge group as $S(U(3)_c\times U(2)_L)$, where the group element $e^{i\theta}\in U(1)_Y$ is identified as $(e^{-i2\theta} \mathbbm{1}_{3\times 3}, e^{i3\theta} \mathbbm{1}_{2\times 2})\in U(3)_c\times U(2)_L$.
\end{enumerate}
  
To elucidate the above results further, we have the following comments. 

It is straightforward to check that all the SM fields are invariant under $\mathbb{Z}_6$, i.e. they satisfy both eqs.~(\ref{eq:quantization_Z6_main_2}) and~(\ref{eq:quantization_Z6_main_1}). For example, one may consider the left-handed quark doublet $Q_L\sim (R_3, R_2, Q_Y)=(\tiny\yng(1),\ \tiny\yng(1),\ 1/6)$. In this example, $Q_L$ transforms as the fundamental $(3+2)$ dimensional representation under $S(U(3)_c\times U(2)_L)$. As a side remark, this choice of quantum numbers is the simplest but not the unique choice.~\footnote{See e.g. Chapter 13 of~\cite{Tung:1985na} for pedagogical explanations on the irreducible representations of $U(N)$ and $SU(N)$ groups. If $Q_L$ is not considered as the fundamental representation of $U(3)_c$ or $U(2)_L$, the smallest possible hypercharge in the entire theory can be smaller than $1/6$ even when the discrete quotient is nontrivial.}

In general, one can also use eqs.~(\ref{eq:quantization_Z6_main_2}) and~(\ref{eq:quantization_Z6_main_1}) to check the examples mentioned in section~\ref{sec:intro} on the four scenarios for all possible UV resonances and the corresponding $\Gamma$. In particular, a particle in representation $\mathcal{R}$ is identified as a $\mathbb{Z}_6$ exotic when at least one of the eqs.~(\ref{eq:quantization_Z6_main_2}) and~(\ref{eq:quantization_Z6_main_1}) is violated. For definiteness, we have only considered the cases where $6 Q_Y$ are integer-valued in section~\ref{sec:intro}. 

When the discrete quotient $\Gamma$ is nontrivial, at least one of the eqs.~(\ref{eq:quantization_Z6_main_2}) and~(\ref{eq:quantization_Z6_main_1}) is satisfied, this implies that $6 Q_Y$ are integer-valued for all the particles in the spectrum. Consequently, discovering particles with fractional $6 Q_Y$ implies that $\Gamma$ must be a trivial group, i.e. $\Gamma=1$. Throughout the paper, we \emph{assume} the $6 Q_Y$ charges for all particles to be rational, such that the gauge group of hypercharge is $U(1)$ rather than $\mathbb{R}^1$. In practice, it is experimentally challenging to distinguish rational versus irrational charges. However, quantum gravity models do prefer all continuous gauge symmetries to be compact~\cite{Banks:2010zn}.

If the $\mathbb{Z}_6$ exotics (i.e. BSM particles in representations $\mathcal{R}$ not invariant under $\mathbb{Z}_6$) are heavier than the weak scale and there is a decoupling limit, one can integrate them out and study their IR imprints using SMEFT. In the rest of the paper, we will analyze the patterns of the effective operators in SMEFT induced by $\mathbb{Z}_6$ exotic particles in more detail. Our study is timely since in recent years there has been significant progress in building the dictionary between the resonances at UV and the effective operators at IR~\cite{deBlas:2017xtg, Li:2022abx, Li:2023cwy, Li:2023pfw, Guedes:2023azv}, and novel methods and tools to perform EFT matching systematically~\cite{Henning:2014wua, Kramer:2019fwz, Gherardi:2020det, Angelescu:2020yzf, Ellis:2020ivx, DasBakshi:2018vni, Cohen:2020qvb, Fuentes-Martin:2020udw, Carmona:2021xtq, Fuentes-Martin:2022jrf, terHoeve:2023pvs, DeAngelis:2023bmd}.

%%%%%%%%%%%%%%%%%%%%%%%%%%%%%%%%%%%%%%%%%%%%%%%%%%%%%%%%%%%%%%%%%%%%%%%%
\subsection{No $\mathbb{Z}_6$ exotics in tree-level UV completions}
%%%%%%%%%%%%%%%%%%%%%%%%%%%%%%%%%%%%%%%%%%%%%%%%%%%%%%%%%%%%%%%%%%%%%%%
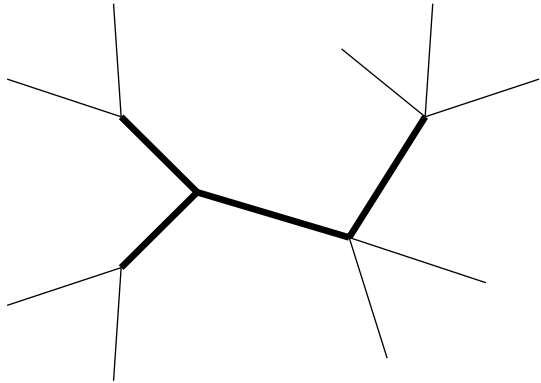
\begin{figure}[t]
\centering
\begin{tikzpicture}
\begin{feynhand}
\vertex[particle] (e1) at (-2.1, 2.5) ;
\vertex[particle] (i9) at (-2.0, 1.0) ;
\vertex[particle] (e2) at (-3.5, 1.5) ;
\vertex[particle] (e3) at (-3.5, -1.5) ;
\vertex[particle] (i10) at (-2.0, -1.0) ;
\vertex[particle] (e4) at (-2.1, -2.5) ;
\vertex[particle] (i11) at (-1.0, 0) ;

\vertex[particle] (e5) at (2.1, 2.5) ;
\vertex[particle] (i13) at (2.0, 1.0) ;
\vertex[particle] (e6) at (3.5, 1.5) ;
\vertex[particle] (e7) at (2.8, -1.2) ;
\vertex[particle] (e8) at (1.5, -2.2) ;
\vertex[particle] (i12) at (1.0, -0.6) ;
\vertex[particle] (e14) at (0.9, 1.9) ;

\propag[plain] (e1) to (i9);
\propag[plain] (e2) to (i9);
\propag[plain] (e3) to (i10);
\propag[plain] (e4) to (i10);
\propag[plain] (e5) to (i13);
\propag[plain] (e6) to (i13);
\propag[plain] (e7) to (i12);
\propag[plain] (e8) to (i12);
\propag[plain] (e14) to (i13);

{\setlength{\feynhandlinesize}{2.5pt}
\propag[plain] (i9) to  (i11);
\propag[plain] (i10) to  (i11);
\propag[plain] (i11) to  (i12);
\propag[plain] (i12) to  (i13);
}
\end{feynhand}
\end{tikzpicture}
\caption{An example of the tree diagrams for generating high-dimensional operators of any mass dimension in SMEFT. The thick internal lines denote the heavy particles responsible for generating the corresponding effective operator, where all the heavy particles must be $\mathbb{Z}_6$ invariant. }
\label{fig:topo_tree}
\end{figure}

It is easy to realize that $\mathbb{Z}_6$ exotics cannot appear in any tree-level UV completions of high-dimensional operators in SMEFT. This result holds for high-dimensional operators of any mass dimensions, and it follows directly from gauge invariance and the fact that $\mathbb{Z}_6$ is a subgroup of $\tilde{G}=SU(3)_c\times SU(2)_L\times U(1)_Y$ acting trivially on all the SM fields.

Let us consider a high-dimensional operator that can be generated by integrating out heavy resonances at tree level. Specifically, we may consider a diagram shown in figure.~\ref{fig:topo_tree}, where all the SM fields are denoted as thin external lines, while heavy resonances are denoted as thick internal lines. 
By cutting any thick internal line (which represents a heavy resonance), the tree diagram in figure.~\ref{fig:topo_tree} gets divided into two sub-diagrams where the thick line being cut also becomes an external one. This is to say that the external lines in each sub-diagram consist of one thick line representing a heavy resonance and some thin lines representing SM fields. 
This implies that, in the sub-diagram, the external heavy resonance can decay into SM particles if its mass is large enough.
Since the $\mathbb{Z}_6$ group already acts trivially on all the SM fields, it must also act trivially on the heavy resonance due to gauge invariance. 
Notice that this result holds for any high-dimensional operators of arbitrary mass dimensions. Of course, the same result applies to the toy model in section~\ref{sec:toymodel} as well. 

We refer the readers to~\cite{Li:2022abx, Li:2023cwy, Li:2023pfw} for concrete examples of heavy particles in tree-level UV completions of SMEFT. As we can see, all the heavy particles are invariant under the $\mathbb{Z}_6$ group.

%%%%%%%%%%%%%%%%%%%%%%%%%%%%%%%%%%%%%%%%%%%%%%%%%%%%%%%%%%%%%%%%%%%%%%%%%%%%%%%%%%%%%%%
\subsection{$\mathbb{Z}_6$ exotics in one-loop matching}
%%%%%%%%%%%%%%%%%%%%%%%%%%%%%%%%%%%%%%%%%%%%%%%%%%%%%%%%%%%%%%%%%%%%%%%%%%%%%%%%%%%%%%%
Since heavy $\mathbb{Z}_6$ exotic particles do not appear in tree-level UV completions of SMEFT, it is strongly motivated to consider loop-level UV completions. We start the discussions by illustrating some general features and then discuss two concrete UV models. For concreteness, we will focus on dimension-six (dim-6) operators in the following. Generalizing the analysis to operators with higher mass dimensions is warranted.

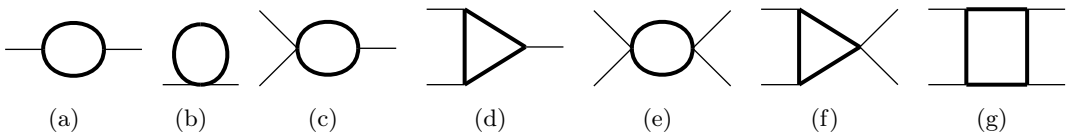
\begin{figure}[t]
\centering
\subfigure[]{
\begin{tikzpicture}
\begin{feynhand}
\vertex[particle] (e1) at (-1.0,0.0) ;
\vertex[particle] (i1) at (-0.5,0) ;
\vertex[particle] (i2) at (0.3,0) ;
\vertex[particle] (e3) at (0.8,0) ;
\propag[plain] (e1) to (i1);
\propag[plain] (i2) to (e3);
{\setlength{\feynhandlinesize}{1.5pt}\propag[plain] (i1) to  [in=90, out=90, looseness=1.5](i2);
\propag[plain] (i1) to  [half right, looseness=1.5](i2);
}
\end{feynhand}
\end{tikzpicture}}
\subfigure[]{
\begin{tikzpicture}
\begin{feynhand}
\vertex[particle] (e1) at (-1.0,0.0) ;
\vertex[particle] (i1) at (-0.5,0) ;
\vertex[particle] (i2) at (-0.5,0.8) ;
\vertex[particle] (e3) at (0,0) ;
\propag[plain] (e1) to (i1);
\propag[plain] (i1) to (e3);
{\setlength{\feynhandlinesize}{1.5pt}\propag[plain] (i1) to [half right, looseness=1.5](i2);
\propag[plain] (i1) to [half left, looseness=1.5](i2);
}
\end{feynhand}
\end{tikzpicture}}
\subfigure[]{
\begin{tikzpicture}
\begin{feynhand}
\vertex[particle] (e1) at (-1.0,0.5) ;
\vertex[particle] (e2) at (-1.,-0.5) ;
\vertex[particle] (i1) at (-0.5,0) ;
\vertex[particle] (i2) at (0.3,0) ;
\vertex[particle] (e3) at (0.8,0) ;
\propag[plain] (e1) to (i1);
\propag[plain] (e2) to (i1);
\propag[plain] (i2) to (e3);
{\setlength{\feynhandlinesize}{1.5pt}\propag[plain] (i1) to  [in=90, out=90, looseness=1.5](i2);
\propag[plain] (i1) to  [half right, looseness=1.5](i2);
}
\end{feynhand}
\end{tikzpicture} 
}
\subfigure[]{
\begin{tikzpicture}
\begin{feynhand}
\vertex[particle] (e1) at (-1.0,0.5) ;
\vertex[particle] (e2) at (-1.,-0.5) ;
\vertex[particle] (i1) at (-0.5,0.5) ;
\vertex[particle] (i2) at (-0.5,-0.5) ;
\vertex[particle] (i3) at (0.3,0.0) ;
\vertex[particle] (e3) at (0.8,0) ;
\propag[plain] (e1) to (i1);
\propag[plain] (e2) to (i2);
\propag[plain] (i3) to (e3);
{\setlength{\feynhandlinesize}{1.5pt}\propag[plain] (i1) to (i2);
\propag[plain] (i2) to (i3);
\propag[plain] (i3) to (i1);
}
\end{feynhand}
\end{tikzpicture} 
}
\subfigure[]{
\begin{tikzpicture}
\begin{feynhand}
\vertex[particle] (e1) at (-1.0,0.5) ;
\vertex[particle] (e2) at (-1.,-0.5) ;
\vertex[particle] (i1) at (-0.5,0) ;
\vertex[particle] (i2) at (0.3,0) ;
\vertex[particle] (e3) at (0.8,0.5) ;
\vertex[particle] (e4) at (0.8,-0.5) ;
\propag[plain] (e1) to (i1);
\propag[plain] (e2) to (i1);
\propag[plain] (i2) to (e3);
\propag[plain] (i2) to (e4);
{\setlength{\feynhandlinesize}{1.5pt}\propag[plain] (i1) to  [in=90, out=90, looseness=1.5](i2);
\propag[plain] (i1) to  [half right, looseness=1.5](i2);
}
\end{feynhand}
\end{tikzpicture} 
}
\subfigure[]{
\begin{tikzpicture}
\begin{feynhand}
\vertex[particle] (e1) at (-1.0,0.5) ;
\vertex[particle] (e2) at (-1.,-0.5) ;
\vertex[particle] (i1) at (-0.5,0.5) ;
\vertex[particle] (i2) at (-0.5,-0.5) ;
\vertex[particle] (i3) at (0.3,0.0) ;
\vertex[particle] (e3) at (0.8,0.5) ;
\vertex[particle] (e4) at (0.8,-0.5) ;
\propag[plain] (e1) to (i1);
\propag[plain] (e2) to (i2);
\propag[plain] (i3) to (e3);
\propag[plain] (i3) to (e4);
{\setlength{\feynhandlinesize}{1.5pt}\propag[plain] (i1) to (i2);
\propag[plain] (i2) to (i3);
\propag[plain] (i3) to (i1);
}
\end{feynhand}
\end{tikzpicture} 
}
\subfigure[]{
\begin{tikzpicture}
\begin{feynhand}
\vertex[particle] (e1) at (-1.0,0.5) ;
\vertex[particle] (e2) at (-1.,-0.5) ;
\vertex[particle] (i1) at (-0.5,0.5) ;
\vertex[particle] (i2) at (-0.5,-0.5) ;
\vertex[particle] (i3) at (0.3,0.5) ;
\vertex[particle] (i4) at (0.3,-0.5) ;
\vertex[particle] (e3) at (0.8,0.5) ;
\vertex[particle] (e4) at (0.8,-0.5) ;
\propag[plain] (e1) to (i1);
\propag[plain] (e2) to (i2);
\propag[plain] (i3) to (e3);
\propag[plain] (i4) to (e4);
{\setlength{\feynhandlinesize}{1.5pt}\propag[plain] (i1) to (i2);
\propag[plain] (i2) to (i4);
\propag[plain] (i4) to (i3);
\propag[plain] (i3) to (i1);
}
\end{feynhand}
\end{tikzpicture} 
}
\caption{All the relevant one-loop topologies for generating the dim-6 operators up to 4 SM fields in SMEFT using the off-shell diagrammatic matching. The thick internal propagators represent $\mathbb{Z}_6$ non-invariant resonances, i.e. the $\mathbb{Z}_6$ exotics. Notice that each vertex must contain two $\mathbb{Z}_6$ exotics due to the requirement of gauge invariance. The two-point (2-pt) diagrams are relevant as they can be converted to the Warsaw basis operators with field redefinition~\cite{Gherardi:2020det, Guedes:2023azv}.
}
\label{fig:topo_loop}
\end{figure}

In figure.~\ref{fig:topo_loop}, we show the corresponding diagrams that are responsible for generating the dim-6 operators in the Green's basis~\cite{Gherardi:2020det, Jiang:2018pbd} up to four SM fields in the matching. We notice that all the heavy particles running in the loops must be non-invariant under the $\mathbb{Z}_6$ group. The reason is that replacing a $\mathbb{Z}_6$ exotic particle with a $\mathbb{Z}_6$-invariant one for any internal propagator necessarily leads to a vertex involving two or three $\mathbb{Z}_6$-invariant particles and one $\mathbb{Z}_6$ exotic particle, where the corresponding three-point (3-pt) or four-point (4-pt) coupling cannot be gauge invariant.

\begin{table}[h]
    \centering
    \begin{tabular}{c|c|c}\hline\hline
        Operators & Scalar $\phi$ & Fermion $\psi$\\
        \hline
        ${\cal O}_{3G} = f^{ABC}G^{A\nu}_\mu G^{B\rho}_\nu G^{C\mu}_\rho$ & $\phi\otimes\bar{\phi}\supset (\mathbf{8},\mathbf{1})$& $\psi\otimes\bar{\psi}\supset (\mathbf{8}, \mathbf{1})$\\
        ${\cal O}_{3W} = \epsilon^{IJK}W^{I\nu}_\mu W^{J\rho}_\nu W^{J\mu}_\rho$ & $\phi\otimes\bar{\phi}\supset (\mathbf{1}, \mathbf{3})$ & $\psi\otimes\bar{\psi}\supset (\mathbf{1},  \mathbf{3})$ \\
        ${\cal O}_{HG} = G^{A}_{\mu\nu}G^{A\mu\nu} H^\dagger H$ &$\phi\otimes\bar{\phi}\supset (\mathbf{8} , \mathbf{1})$ & N.A.\\
        ${\cal O}_{HW} = W^{I}_{\mu\nu}W^{I\mu\nu}H^\dagger H$ &$\phi\otimes\bar{\phi}\supset (\mathbf{1},  \mathbf{3}) $ & N.A.\\
        ${\cal O}_{HB} = B_{\mu\nu}B^{\mu\nu}H^\dagger H$ & $Y_{\phi}\neq 0$ & N.A.\\
        ${\cal O}_{HWB} = W^{I}_{\mu\nu}B^{\mu\nu}H^\dagger\sigma^I H$ & $\phi\otimes\bar{\phi}\supset (\mathbf{1}, \mathbf{3}),\ Y_\phi\neq 0$ & N.A.\\
        \hline\hline
    \end{tabular}
    \caption{The necessary conditions for generating the corresponding loop-level generated dim-6 operators in the Warsaw basis with a \emph{single} heavy scalar field $\phi$ or a fermion field $\psi$. {{The quantum number of the gluon is denoted by $(\mathbf{8} , \mathbf{1})$, while that for the $W$ boson is $(\mathbf{1}, \mathbf{3})$.}} `N.A.' means the corresponding operator cannot be generated at the one-loop level with only renormalizable interactions in the UV models. This motivates us to consider a less minimal model with two $\mathbb{Z}_6$ exotic heavy fermions. }
    \label{tab:oneZ6B}
\end{table}

For definiteness, we will consider minimal models that contain only one $\mathbb{Z}_6$ exotic heavy particle. In such a scenario, more constraints can be deduced on the couplings between the $\mathbb{Z}_6$ exotic and SM particles.
According to the previous discussion, the $\mathbb{Z}_6$ exotic heavy particle must appear in conjugate pairs in each vertex in the one-loop diagrams, and the total hypercharge of the conjugate pair is zero. This implies that the total $U(1)_Y$ charge of the SM fields in the vertex must also vanish. 
For the 3-pt vertices in figure.~\ref{fig:topo_loop}, the SM fields must be the gauge bosons of $SU(3)_c\times SU(2)_L \times U(1)_Y$, whose field strengths $\{ G^{A}_{\mu\nu}, W^{I}_{\mu\nu}, B_{\mu\nu} \}$ can collectively be denoted as $F_{\mu\nu}$. For the 4-pt vertices in figure.~\ref{fig:topo_loop}, the two SM fields must be $H^\dagger H$, where $H$ is the SM Higgs doublet. Consequently, models with one $\mathbb{Z}_6$ exotic heavy particle contribute to the dim-6 Green's basis operators of the types $F^3$, $F^2 H^2$, $H^2FD^2$, $(DF)^2$, $H^4D^2$ and $|H|^6$, among which $H^2FD^2$, $(DF)^2$ and some of the operators of $H^4D^2$ will be converted to the operators of the types $F^2 H^2$, $H^3f^2$, $H^2f^2 D$, $|H|^6$ and $f^4$ when reducing to the Warsaw basis using field redefinitions~\cite{Gherardi:2020det, Jiang:2018pbd}, where $f$, $H$, $F$ and $D$ represent the SM fermions, the Higgs field, the gauge field strengths, and the covariant derivative, respectively. 

For our later discussion, we list in table.~\ref{tab:oneZ6B} the dim-6 operators that can only be generated at loop level and the possible UV completions with one scalar or one fermion.
Obviously, for a $\mathbbm{Z}_6$ exotic that contribute to Green's basis operators containing $G^A_{\mu\nu}$, it must be charged under $SU(3)_c$, and similar arguments apply to the gauge field strength tensors $W^I_{\mu\nu}$ and $B_{\mu\nu}$ for $SU(2)_L$ and $U(1)_Y$ gauge groups, respectively.

In the subsequent sections, we will focus on two classes of benchmark models, perform the one-loop matching, and solve for the quantum numbers of the $\mathbb{Z}_6$ exotics in terms of Wilson coefficients. 
As will be discussed in detail in section~\ref{sec:pheno}, this information is crucial for testing these models with future measurements.

\subsection{Minimal model 1: scalar extension}
We consider the simplest extension with only one scalar field, denoted as $\phi$, transforming as a $\mathbb{Z}_6$ exotic in a general representation $(R_3, R_2, Y_\phi)$ under $\tilde{G}=SU(3)_c\times SU(2)_L \times U(1)_Y$. 
Since $\phi$ is not invariant under the $\mathbb{Z}_6$ subgroup of $\tilde{G}$, at least one of the eqs.~(\ref{eq:quantization_Z6_main_1}) and~(\ref{eq:quantization_Z6_main_2}) is not satisfied. For example, a $\mathbb{Z}_6$ exotic $\phi$ can be $(\tiny\yng(1)\ , \tiny\yng(1)\ , 1/3)$ under $\tilde{G}=SU(3)_c\times SU(2)_L \times U(1)_Y$.

The Lagrangian for this minimal model contains the following terms
\bea
    {\cal L}_{\phi}\supset && (D_\mu\phi^\dagger)(D^\mu \phi) - M^2 \phi^\dagger \phi-\lambda_{\mathbf{3}}(H^\dagger\sigma^I H)(\phi^\dagger T^I\phi)-\lambda_{\mathbf{1}}(H^\dagger H) (\phi^\dagger\phi)\ ,\label{eq:Lscalar}
\eea
where the SM Lagrangian and self-interacting quartic terms of $\phi$ are not explicitly shown, $\sigma^{I}$ and $T^{I}$ ($I=1,2,3$) are respectively the usual Pauli matrices and the generators for the $SU(2)_L$ representation $R_2$. 
In particular, we notice that there is no 3-pt vertex with two Higgs bosons and one $\phi$, since this interaction cannot be gauge invariant. Furthermore, there is no 3-pt vertex with two $\phi$'s and one Higgs boson, because $H$ is in a half-integer spin representation of $SU(2)_L$, while there are only integer-spin representations of $SU(2)_L$ in the decomposition of $\bar{R}_2\otimes {R}_2$. 
In principle, there can be 4-pt vertices with three $\phi$ and one $H$, but those terms do not contribute to one-loop matching.

Based on the Lagrangian in eq.~\eqref{eq:Lscalar}, we leave the quantum numbers of $\phi$ as undetermined variables and perform the one-loop matching using the method of covariant derivative expansion (CDE)~\cite{Henning:2014wua}. The Wilson coefficients for the bosonic sector operators in Warsaw basis are obtained as follows:
\begin{eqnarray}
    c_{3G}&=& \frac{g_3^3}{(4\pi)^2 180 M^2}\ \mu(R_3)\ d(R_2)\ ,
    \quad c_{3W}= \frac{g_2^3}{(4\pi)^2 180 M^2}\ \mu(R_2)\ d(R_3) \ ,\\
    c_{HG}&=& \frac{g_3^2 \lambda_{\mathbf{1}}}{(4\pi)^212 M^2}\ \mu(R_3)\ d(R_2)\ ,
    \quad  c_{HW}= \frac{g_2^2 \lambda_{\mathbf{1}}}{(4\pi)^2 12 M^2}\ \mu(R_2)\ d(R_3)\ ,\\  
    c_{HB}&=& \frac{g_1^2 Y_\phi^2 \lambda_{\mathbf{1}}}{(4\pi)^2 12 M^2}\ d(R_2)\ d(R_3)\ , 
    \quad c_{HWB}= \frac{g_1 g_2 Y_\phi \lambda_{\mathbf{3}}}{(4\pi)^2 6 M^2}\ \mu(R_2)\ d(R_3)\ ,\\
    c_{H\Box}&=&-\frac{1}{(4\pi)^2 12M^2 }\left[ d(R_2)d(R_3)\left(\lambda_{\mathbf{1}}^2+\frac{g_1^4 Y_\phi^2}{20}\right)+\mu(R_2)d(R_3)\left(\frac{3g_2^2}{80}-\lambda_{\mathbf{3}}^2\right)\right],\\
    c_{HD}&=&-\frac{1}{(4\pi)^2 3 M^2 }\lambda^2_{\mathbf{3}}\mu(R_2)d(R_3)-\frac{g_1^4}{(4\pi)^2 60 M^2}{Y_\phi ^2 d(R_3)d(R_2)}\label{eq:ScHD},
\end{eqnarray}
where $d(R_i)$ and $\mu(R_i)$ are the dimensions and the Dynkin indices for the corresponding representations under $SU(3)_c$ or $SU(2)_L$, whose values can be calculated using the formulae in appendix~\ref{app:derivation}. The gauge couplings of $SU(3)_c, SU(2)_L, U(1)_Y$ are denoted as $g_{3,2,1}$, and $c_{H\Box}$ and $c_{HD}$ are the Wilson coefficients for the operators ${\cal O}_{H\Box}=|H|^2\Box |H|^2$ and ${\cal O}_{HD}=|H^\dagger D_\mu H|^2$, respectively.  We also check the above Wilson coefficients using \texttt{matchete}~\cite{Fuentes-Martin:2022jrf}. 

To facilitate the following discussion, we also present the Wilson coefficients for two additional four-fermion operators
${\cal O}^{prst}_{ee}=(\bar{e}_{Rp}\gamma^\mu e_{Rr})(\bar{e}_{Rs}\gamma_{\mu}e_{Rt})$
and ${\cal O}^{prst}_{ll}=(\bar{l}_p\gamma^\mu l_r)(\bar{l}_s\gamma_{\mu}l_t)$:
\begin{eqnarray}
      c_{ee}&\equiv& c_{ee}^{pppp}= -\frac{g_1^4 Y^2_\phi}{(4\pi)^2 60 M^2}d(R_2)d(R_3), \\
       c_{ll}&\equiv& c_{ll}^{pppp}=-\frac{1}{(4\pi)^2 240 M^2}\left[g_2^4\mu(R_2)d(R_3)+g_1^4Y^2_{\phi}d(R_3)d(R_2)\right],
\end{eqnarray}
where four $p$ in the superscripts indicate that the four flavor indices of the fermions are the same.  ${\cal O}^{prst}_{ee}$ comes from the conversion of the Green's basis operator $(\partial_\mu B^{\mu\nu})^2$, while ${\cal O}^{prst}_{ll}$ gets contribution from the conversion from the operators $(\partial_\mu B^{\mu\nu})^2$ and $(D_\mu W^I_{\mu\nu})^2$ in the Green's basis. This also explains why each term is proportional to gauge couplings to the fourth power, two from the direct matching of the 2-pt amplitude as indicated in the first diagram in figure.~\ref{fig:topo_loop}, and another two from the conversion of $D_\mu F_{\mu\nu}$ using equations of motion. 
The same observation also applies to ${\cal O}_{HD}$, where the second term in eq.~\eqref{eq:ScHD} is from $(\partial_\mu B^{\mu\nu})^2$. 

From these observations, one can solve the quantum numbers of $\phi$ in terms of Wilson coefficients and gauge couplings:
\begin{eqnarray}
Y_\phi^2 &=& \frac{g_2 c_{HWB}^2}{15 g_1^2 c_{3W}(c_{ee}-c_{HD})}\ , \label{eq:scalar_Y} \\
\frac{\mu(R_3)}{d(R_3)} &=& \frac{g_2 c_{HG}c_{HWB}^2}{15 g_3^2 c_{HB} c_{3W}(c_{ee}-c_{HD})}\ , \label{eq:scalar_3} \\
\frac{\mu(R_2)}{d(R_2)} &=& \frac{ c_{HW}c_{HWB}^2}{15 g_2 c_{HB} c_{3W}(c_{ee}-c_{HD})}\ . \label{eq:scalar_2}
\end{eqnarray}
These solutions are valid provided that $\phi$ transforms non-trivially under all the three gauge groups. 
Otherwise, if $\phi$ is a singlet under $SU(2)_L$ for example, then $c_{3W}$ is zero and it should not appear in the denominators of the above formulae. 
Depending on the additional assumptions on the gauge representations of $\phi$, one can solve the quantum numbers using different combinations of Wilson coefficients. For instance, if $\phi$ is only charged non-trivially under $SU(3)_c$, then the Wilson coefficients $c_{3W}$, $c_{HW}$, $c_{HB}$, $c_{HWB}$, $c_{HD}$ are all zero, and one can construct the solution $\mu(R_3)/d(R_3)$ as follows:
\begin{eqnarray}
    \frac{\mu(R_3)}{d(R_3)} = -\frac{c_{HG}^2}{15g_3 c_{H\Box}c_{3G}}.\label{eq:R3nontrivial}
\end{eqnarray}
Depending on the assumptions on the quantum numbers of $\phi$, we summarize the formulae for the solutions in table.~\ref{tab:Ssolutions}.

\begin{table}
\centering
\renewcommand{\arraystretch}{1.5}
\begin{tabular}{ |c|c| } 
 \hline
 Representation & Solution  \\
 \hline
 $\phi\ (\cdot, \cdot, Y_\phi)$ & $Y_\phi^2=\frac{4c_{HB}^2}{5(4c_{H\Box}-c_{HD})c_{HD}}$  \\ 
 \hline
  $\phi (R_3,\cdot, 0)$ & $\frac{\mu(R_3)}{d(R_3)} = -\frac{c_{HG}^2}{15g_3 c_{H\Box}c_{3G}}$  \\ 
 \hline
    $\phi (\cdot,R_2, 0)$ &$\frac{\mu(R_2)}{d(R_2)} = \frac{80 c^2_{HW}c_{ll}}{225c_{3W}^2\left[g_2^2\left(4c_{H\Box}+c_{HD}\right)-3c_{ll}\right]}$ \\
 \hline
   {$\phi\ (R_3, \cdot, Y_\phi)$} & $Y_\phi^2 = \frac{4 {{g_3^2}} c_{HG}^2c_{HD}}{45g_1^4c^2_{3G}(4c_{H\Box}-c_{HD})}\quad \frac{\mu(R_3)}{d(R_3)} = \frac{4 c^3_{HG}c_{HD}}{45g_1^2 c^2_{3G}c_{HB}(4c_{H\Box}-c_{HD})}$
   %$Y_\phi^2 = -\frac{4g_3 c_{HB}c_{HG}}{15g_1^2(4c_{HD}-c_{HD})c_{3G}}\quad 
   % \frac{\mu(R_3)}{d(R_3)} =-\frac{4g_3 c_{HG}^2}{15(4c_{HD}-c_{HD})c_{3G}}$
   \\
 \hline
    {$\phi\ (\cdot, R_2, Y_\phi)$} & $Y_\phi^2 = \frac{g_2 c^2_{HWB}}{15 g_1^2 c_{3W}(c_{ee}-c_{HD})}\quad \frac{\mu(R_2)}{d(R_2)} = - \frac{g_1^2 c_{HWB}^2}{5 g^2_2 c_{ee}(c_{ee}-c_{HD})}$\\
  \hline
   \multirow{2}{5em}{$\phi\ (R_3, R_2, 0)$} & $
   %\frac{\mu(R_3)}{d(R_3)} = -\frac{g_3^2 c^2_{HG}}{c_{3G}}\left[ 15 g_3^2c_{H\Box}+\frac{45}{16}c_{3W}\left(3g_2^2-\frac{g_2^4 c_{HD}}{c_{ll}}\right)\right]^{-1}$
   \frac{\mu(R_3)}{d(R_3)} =-\frac{c_{HG}^2}{15g_3 c_{3G}(c_{H\Box}-3c_{ll}/(4g_2^2)+c_{HD}/4)}$
   \\
   & $\frac{\mu(R_2)}{d(R_2)} = -\frac{c_{HW}^2}{15g_2 c_{3W}(c_{H\Box}-3c_{ll}/(4g_2^2)+c_{HD}/4)}$
%         \frac{\mu(R_2)}{d(R_2)} = -\frac{c_{HW}^2}{c_{3W}}\left[15 g_2 c_{H\Box}+\frac{45}{16}\left(3-g_2^2\frac{c_{HD}}{c_{ll}}\right)\right]^{-1}
   \\
   \hline
    
 \hline
\end{tabular}\caption{Summary of the solutions to the quantum numbers of $\phi$. }\label{tab:Ssolutions}
\end{table}

\subsection{Minimal model 2: fermion extension}

We consider the second minimal model with two Dirac fermions which are not invariant under the $\mathbb{Z}_6$ group. These two fermions are denoted as $\psi_{1,2}$.
For models with a single $\mathbb{Z}_6$ exotic Dirac fermion, no renormalizable interaction term can be written down in the Lagrangian other than the gauge couplings, and we will comment on their SMEFT implications at the end of this section. 

For the model with two $\mathbb{Z}_6$ exotic Dirac fermions $\psi_{1,2}$, one can write down the Yukawa coupling with the SM Higgs doublet when the quantum numbers of the fermions satisfy the constraints $\overline{\psi}_1\otimes \psi_2 \supset (\mathbf{1}, \mathbf{2})$ under $SU(3)_c\times SU(2)_L$ and $Y_{\psi_1}=Y_{\psi_2}+\frac{1}{2}$. This implies that the $SU(3)$ quantum numbers of $\psi_1$ and $\psi_2$ are the same, and their $SU(2)$ quantum numbers differ by isospin $1/2$.~\footnote{Another choice is that $Y_{\psi_1}=Y_{\psi_2}-\frac{1}{2}$, the corresponding Yukawa coupling becomes $-\lambda_{\psi} (\overline{\psi}_1 \tilde{H} \psi_2+\text{h.c.})$, where $\tilde{H}=i\sigma_2 H$. } Let us consider the following Lagrangian 
\begin{eqnarray}
    {\cal L}_{\psi}\supset \overline{\psi}_1(i\slashed{D}-M_1)\psi_1+\overline{\psi}_2(i\slashed{D}-M_2)\psi_2- \lambda_{\psi} (\overline{\psi}_{1i}H_n\psi_{2j}\Gamma_{ij}^n+\text{h.c.})\ ,\label{eq:Lfermion}
\end{eqnarray}
where $H$ is the SM Higgs field, and for simplicity we neglect the parity-violating term $\bar{\psi}_1 i\gamma_5\psi_2 H$ in this work. Without loss of generality, we consider $\psi_1$ in the spin $R_2$ representation of $SU(2)_L$ and $\psi_2$ in the spin $(R_2+1/2)$ representation, then $\Gamma_{ij}^n$ is nothing but the Clebsch-Gordon coefficient for which we adopt the following normalization condition
\begin{eqnarray}
   \sum_{ij} \Gamma_{ij}^n  (\Gamma_{ij}^m)^* = \delta^{mn} .
\end{eqnarray}
Similar to the scalar model, we perform one-loop matching using the universal one-loop effective action for heavy fermion fields~\cite{Ellis:2020ivx}, and we take the assumption $M_1 = M_2 = M$ for simplicity. The Wilson coefficients are
\begin{eqnarray}
    c_{3G} &=& -\frac{g_3^3}{(4\pi)^2 45 M^2 }\mu(R_3)(d(R_2)+1/2),\\
    c_{3W} &=& -\frac{g_2^3}{(4\pi)^2 90 M^2 }d(R_3)(\mu(R_2)+\mu(R_2+1/2)), \\
     c_{HG}&=& \frac{-g_3^2 |\lambda_{\psi}|^2}{(4\pi)^2 3 M^2}\ \mu(R_3),\quad \\
     c_{HW}&=& \frac{-g_2^2|\lambda_{\psi}|^2d(R_3)}{(4\pi)^2 M^2}\ F(R_2),\\  
    c_{HB}&=& \frac{-g_1^2 |\lambda_{\psi}|^2d(R_3)}{(4\pi)^2 M^2}\ \left[ \frac{(Y_{\psi_2}+1/2)^2+Y_{\psi_2}^2}{3}-\frac{1}{40}\right], \\
%    c_{HWB}&=& \frac{-g_1 g_2 |\lambda_{\psi}|^2d(R_3)}{(4\pi)^2 M^2}\ \left[ \frac{2(Y_\phi c_4(R_2)+(Y_\phi+1/2)c_{3}(R_2))}{3}+\frac{c_4(R_2)-c_3(R_2)-1}{40}\right],\\
    c_{quqd1} &\equiv& c^{3333}_{quqd1} =\frac{ |\lambda_{\psi}|^2}{(4\pi)^25 M^2} d(R_3){y}_{u}^{33}y^{33}_d,
   % c_{H\Box}&=&-\frac{1}{(4\pi)^2 12M^2 }\left[ d(R_2)d(R_3)\left(\lambda_{\mathbf{1}}^2+\frac{g_1^4 Y_\phi^2}{20}\right)+\mu(R_2)d(R_3)\left(\frac{3g_2^2}{80}-\lambda_3^2\right)\right],\\
   % c_{HD}&=&-\frac{1}{(4\pi)^2 3 M^2 }\lambda^2_{\mathbf{3}}\mu(R_2)d(R_3)-\frac{g_1^4}{(4\pi)^2 60 M^2}{Y_\phi ^2 d(R_3)d(R_2)}\label{eq:FcHD},
\end{eqnarray}
where $c^{prst}_{quqd1}$ is the Wilson coefficient for the operator $\mathcal{O}^{prst}_{quqd1} = (\bar{q}^i_pu_r)\epsilon^{ij} (\bar{q}^j_rd_t)$, and $y_e$ and $y_d$ are the Yukawa matrices for the lepton and down-type quark sectors. 
We choose the flavor indices to be the third-generation ones, since the corresponding Yukawa couplings are the largest ones. 
The group theoretical constant $F(R_2)$ is defined by the following equations:
\begin{eqnarray}
    &&F(R_2) = \left[ \frac{c_5(R_2)+c_{6}(R_2)}{3}+\frac{c_4(R_2)-c_3(R_2)}{40}\right]\ ,\\
    &&\Gamma^n_{ij}(\Gamma^q_{kj})^*T^I_{1ki} = c_3(R_2)\tau^{I}_{qn},\quad (\Gamma^m_{jk})^*\Gamma^p_{jl}T_{2lk}^I = c_4(R_2)\tau^{I}_{mp},\\
    &&\frac{1}{2}\Gamma^n_{ij}(\Gamma^q_{kj})^*\{T_1^I,T_1^J\}_{ki} = c_5(R_2)\delta^{IJ}\delta^{nq},\quad \frac{1}{2}(\Gamma^m_{jk})^*\Gamma^p_{jl}\{T_{2}^I,T_2^J\}_{lk} = c_6(R_2)\delta^{IJ}\delta^{mp},\nonumber \\
\end{eqnarray}
where $T_1$ and $T_2$ are the $SU(2)_L$ generators for $\psi_1$ and $\psi_2$ respectively, and $\tau^I = \sigma^I/2$ are the generators of the fundamental representation of $SU(2)_L$, and the curly brackets represent anti-commutators. We illustrate the numerical value of these constants in table.~\ref{tab:cs}. In practice, we obtain the numerical values of the generator matrices using the package \texttt{GroupMath}~\cite{Fonseca:2020vke}.
Being different from the scalar model, the fermion model generates the four fermion operators containing the SM fermion fields of opposite chiralities, i.e. $(\bar{q}u)\epsilon(\bar{q}d)$. These operators come from the conversion of the Green's basis operator $(D^2H)^\dagger (D^2 H)$ using equations of motion.

\begin{table}
\renewcommand{\arraystretch}{1.5}
\setlength{\tabcolsep}{7pt}
\begin{tabular}{ |c|c|c|c|c|c|c|c|c|c|c|c|c|c|c| } 
 \hline
 $d(R_2)$ & 1 & 2 & 3 & 4 & 5 & 6 & 7 & 8 & 9 & 10 & 11 & 12 & 13 & 14 \\
 \hline
 $c_3(R_2)$& 0& $-\frac{1}{3}$& $\frac{-2}{3}$& -1& $\frac{-4}{3}$& $\frac{-5}{3}$& -2& $\frac{-7}{3}$& $\frac{-8}{3}$& -3& $\frac{-10}{3}$& $\frac{-11}{3}$& -4& $\frac{-13}{3}$\\
 \hline
 $c_4(R_2)$& -1& $\frac{-4}{3}$& $\frac{-5}{3}$& -2& $\frac{-7}{3}$& $\frac{-8}{3}$& -3& $\frac{-10}{3}$& $\frac{-11}{3}$& -4& $\frac{-13}{3}$& $\frac{-14}{3}$& -5& $\frac{-16}{3}$\\
 \hline
 $c_5(R_2)$ &0& $\frac{1}{4}$ & $\frac{2}{3}$ & $\frac{5}{4}$ & 2& $\frac{35}{12}$ & 4& $\frac{21}{4}$ & $\frac{20}{3}$ & $\frac{33}{4}$ & 10& $\frac{143}{12}$ & 14& $\frac{65}{4}$  \\
 \hline
 $c_6(R_2)$ & $\frac{1}{4}$& $\frac{2}{3}$& $\frac{5}{4}$& 2& $\frac{35}{12}$& 4& $\frac{21}{4}$& $\frac{20}{3}$& $\frac{33}{4}$& 10& $\frac{143}{12}$& 14& $\frac{65}{4}$& $\frac{56}{3}$ \\
 \hline
\end{tabular}\caption{Numerical values for the group structure constants $c_{3-6}$.}\label{tab:cs}
\end{table}

One can solve for the quantum numbers of $\psi_1$ and $\psi_2$ with the Wilson coefficients and the couplings in the SM as follows:
\begin{eqnarray}
   F(R_2) &=& -\frac{y^{33}_{u}y^{33}_dc_{HW}}{5 g_2^2c_{quqd1}}, \label{eq:F2} \\
   \frac{\mu(R_3)}{d(R_3)} &=& -\frac{3c_{HG}y^{33}_{u}y^{33}_d}{5g_3^2 c_{quqd1}}, \label{eq:psi3} \\
   (Y_{\psi_2}+1/2)^2+Y_{\psi_2}^2 &=& \frac{3}{40}-\frac{3y^{33}_{u}y^{33}_dc_{HB}}{5g_2^2c_{quqd1}}.\label{eq:Ypsi}
\end{eqnarray}
The above formulae do not have singularities as $c_{quqd1}$ cannot be zero as long as $\lambda_\psi\neq 0$, so there is no need for a table akin to the table.~\ref{tab:Ssolutions}.
Furthermore, one can also solve for $R_2$ using the Wilson coefficients of the bosonic operators only, and we define the relevant group theoretical constant $G(R_2)$ as follows:
\begin{eqnarray}
    G(R_2) \equiv  \frac{6 F(R_2)(d(R_2)+1/2)}{\mu(R_2)+\mu(R_2+1/2)} =\frac{g_2c_{HW}c_{3G}}{ g_3c_{HG}c_{3W}}\ .
\end{eqnarray}

Lastly, we comment on the model with a single $\mathbb{Z}_6$ exotic fermion. Since there is no interaction term at the renormalizable level, the only Green's basis operators that can be generated are of the types of $F^3$ and $(DF)^2$. Therefore we have the following formulae for Wilson coefficients in the Green's basis from the matching:
\begin{eqnarray}
    c_{3G} &=& -\frac{g_3^3}{(4\pi)^2 90M^2}\mu(R_3)d(R_2)\ ,\quad c_{3W} =-\frac{g_2^3}{(4\pi)^2 90M^2}\mu(R_2)d(R_3)\ ,
    \\
    c_{2G} &=&-\frac{2g_3^3}{(4\pi)^2 15M^2}\mu(R_3)d(R_2)\ ,\quad c_{2W} =-\frac{2g_2^3}{(4\pi)^2 15M^2}\mu(R_2)d(R_3)\ ,
    \\
    c_{2B} &=&-\frac{2g_1^3}{(4\pi)^2 15M^2}d(R_3)d(R_2)Y_{\psi}^2\ ,
\end{eqnarray}
where the Wilson coefficients $c_{2F}$ are for the operators $(D_\mu F_{\mu\nu})^2$, which can be converted to four fermion operators and $H^4D^2$ operators in the Warsaw basis using equations of motion. It turns out that one cannot obtain a set of solutions for the three quantum numbers separately. In this case, one can only take the ratios of these Wilson coefficients to eliminate the heavy scale $M$.

%%%%%%%%%%%%%%%%%%%%%%%%%%%%%%%%%%%%%%%%%%%%%%%%%%%%%%%%%%%%%%%%%%%%%%%%%%
\section{General Strategy for probing $\mathbb{Z}_6$ exotics with SMEFT}\label{sec:pheno}
%%%%%%%%%%%%%%%%%%%%%%%%%%%%%%%%%%%%%%%%%%%%%%%%%%%%%%%%%%%%%%%%%%%%%%%%%%

In this section, we delineate a comprehensive strategy aimed at probing the aforementioned $\mathbb{Z}_6$ exotic models through an analysis of the Wilson coefficients, whose values will be determined in forthcoming experiments in the future. In general, the information on UV physics is encoded in the correlations between different Wilson coefficients of the operators in SMEFT. Our major point is to demonstrate that the quantized (i.e. discrete) nature of $SU(3)_c$ and $SU(2)_L$ quantum numbers can serve as pivotal correlations among Wilson coefficients for probing a class of $\mathbb{Z}_6$ exotic models.  

In the first step, one can determine whether the $\mathbb{Z}_6$ exotic particle in the hypothesis test transforms trivially under some SM gauge group factors by finding the exceedingly small Wilson coefficients from the measurements. Taking the scalar model as an example, if $c_{3G}$, $c_{HG}$, $c_{HB}$, $c_{HWB}$ and $c_{ee}$ are measured to be exceedingly small but $c_{3W}$, $c_{HW}$, $c_{HD}$, $c_{H\Box}$ and $c_{ll}$ are measured to be nonzero, it indicates that the UV $\mathbb{Z}_6$ exotic particle may only be charged non-trivially under $SU(2)_L$.  

\begin{figure}[t]
    \centering
    \includegraphics[width = 7cm]{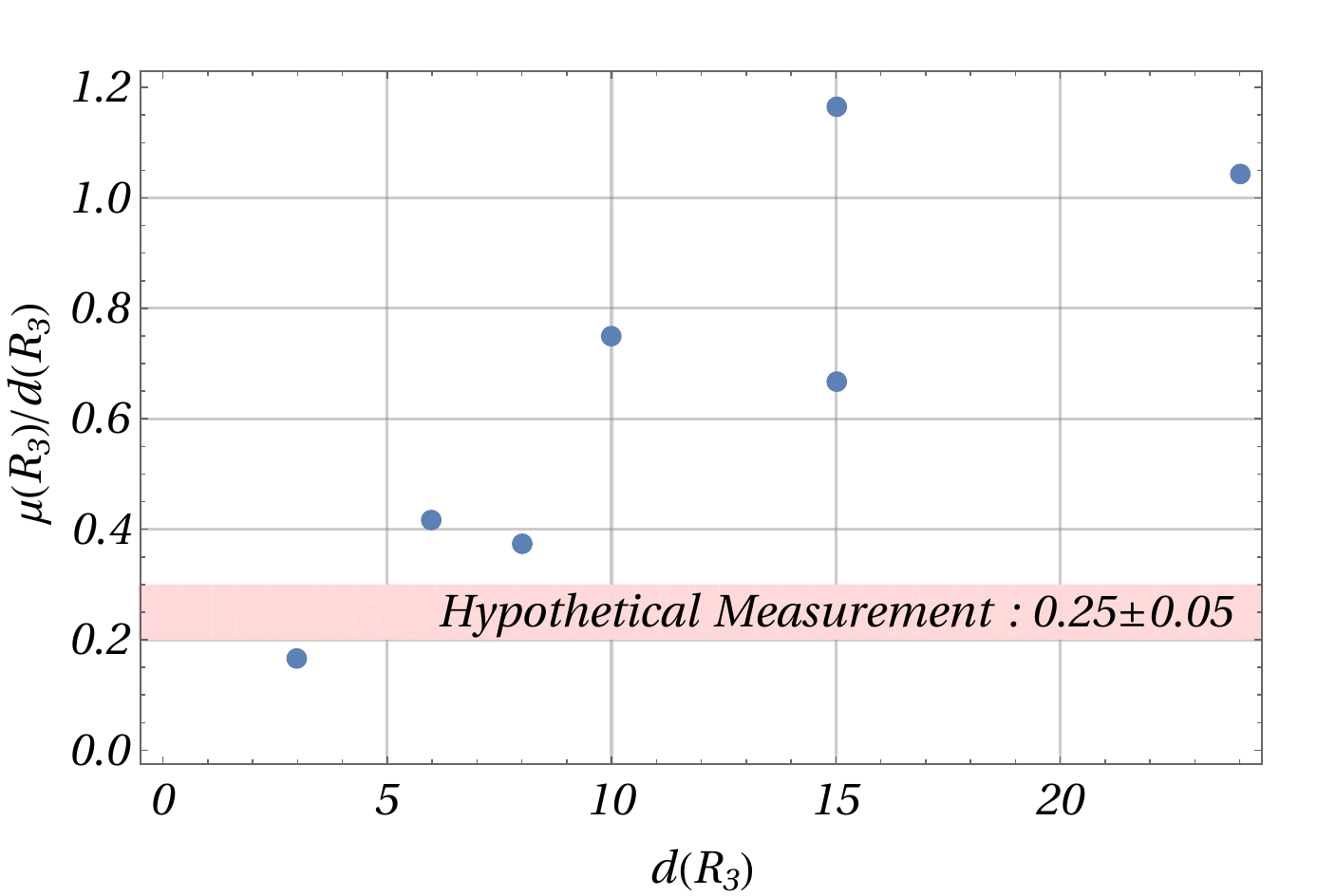}
    \includegraphics[width = 7cm]{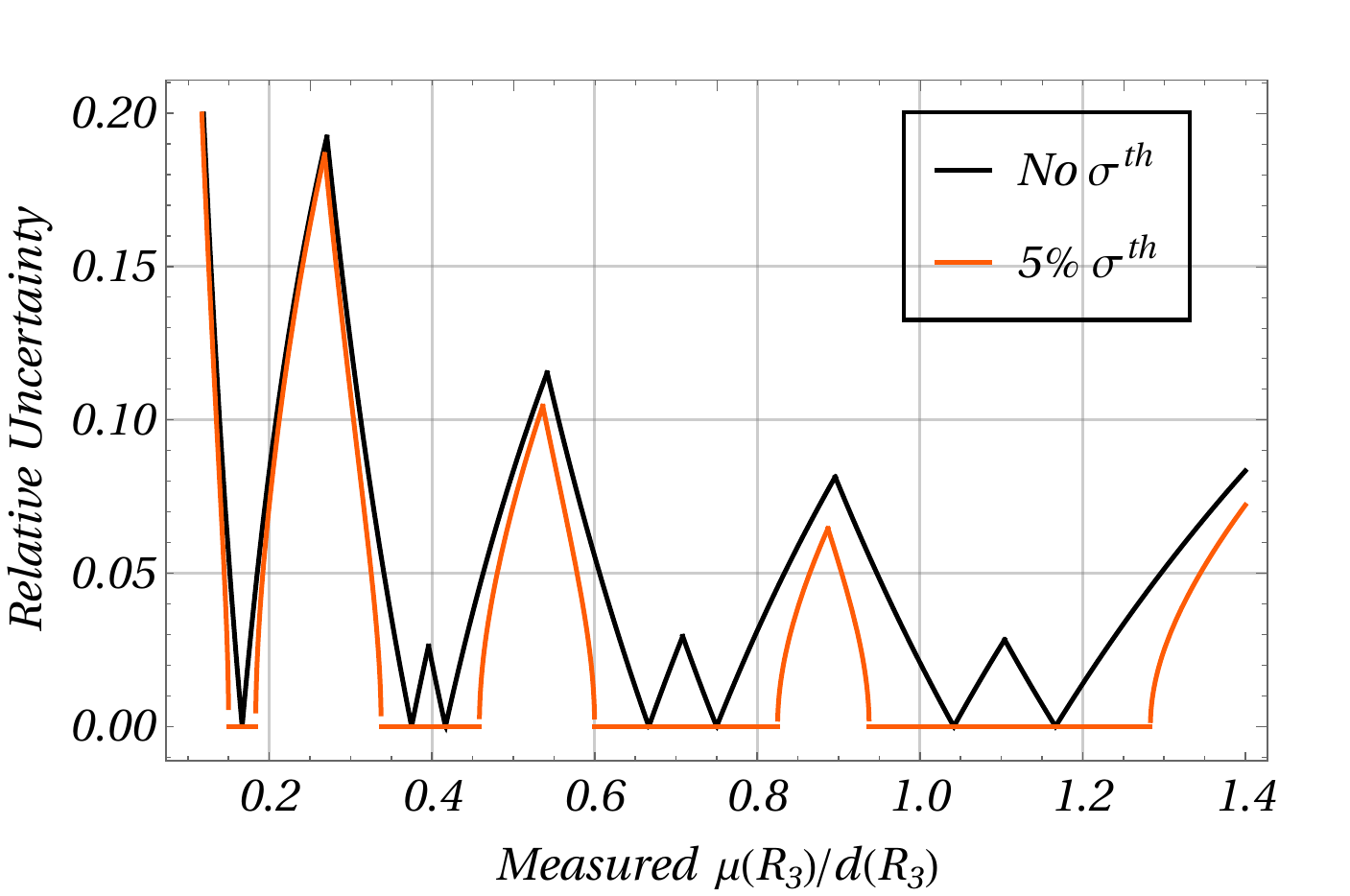}
    \includegraphics[width = 7cm]{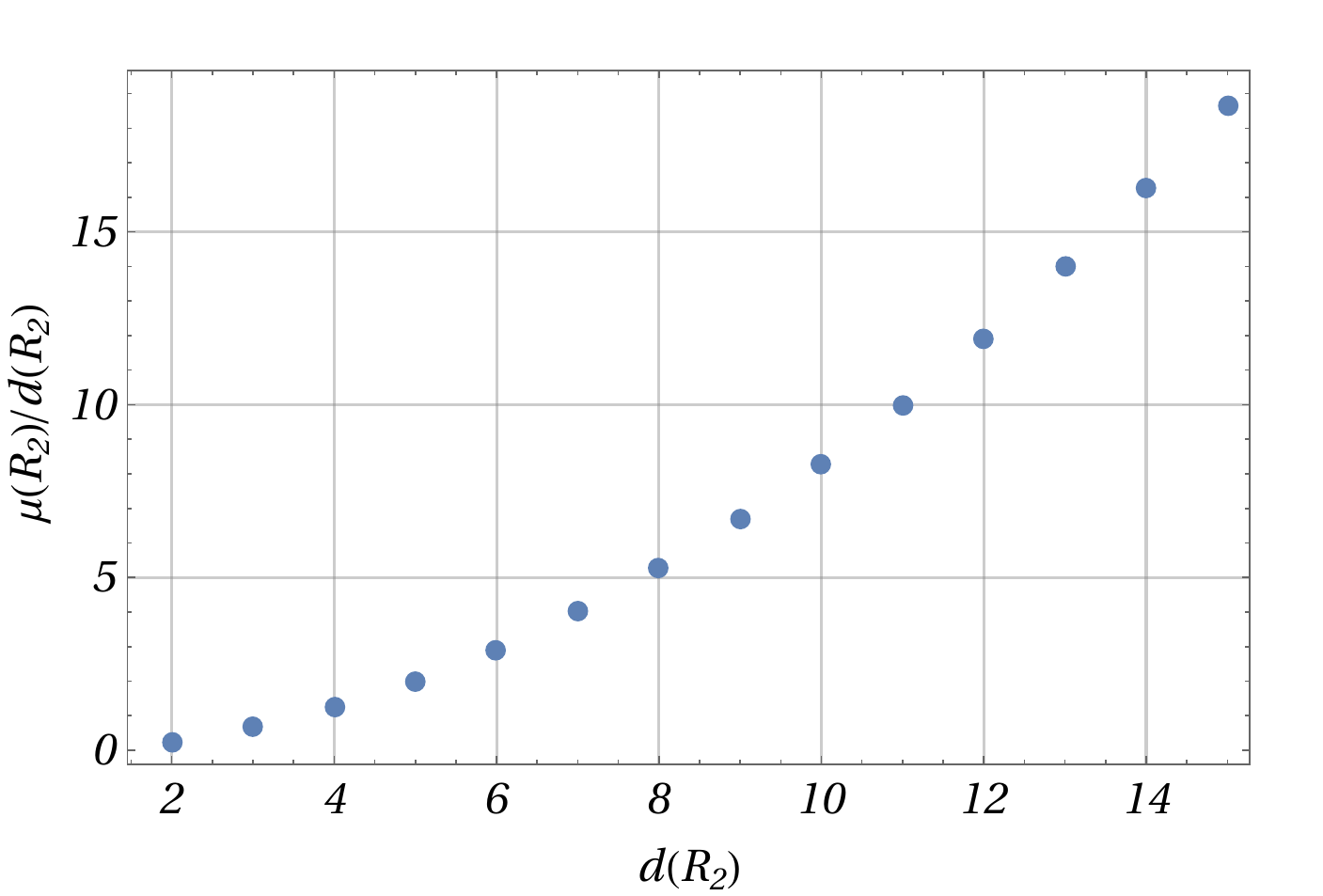}
    \includegraphics[width = 7cm]{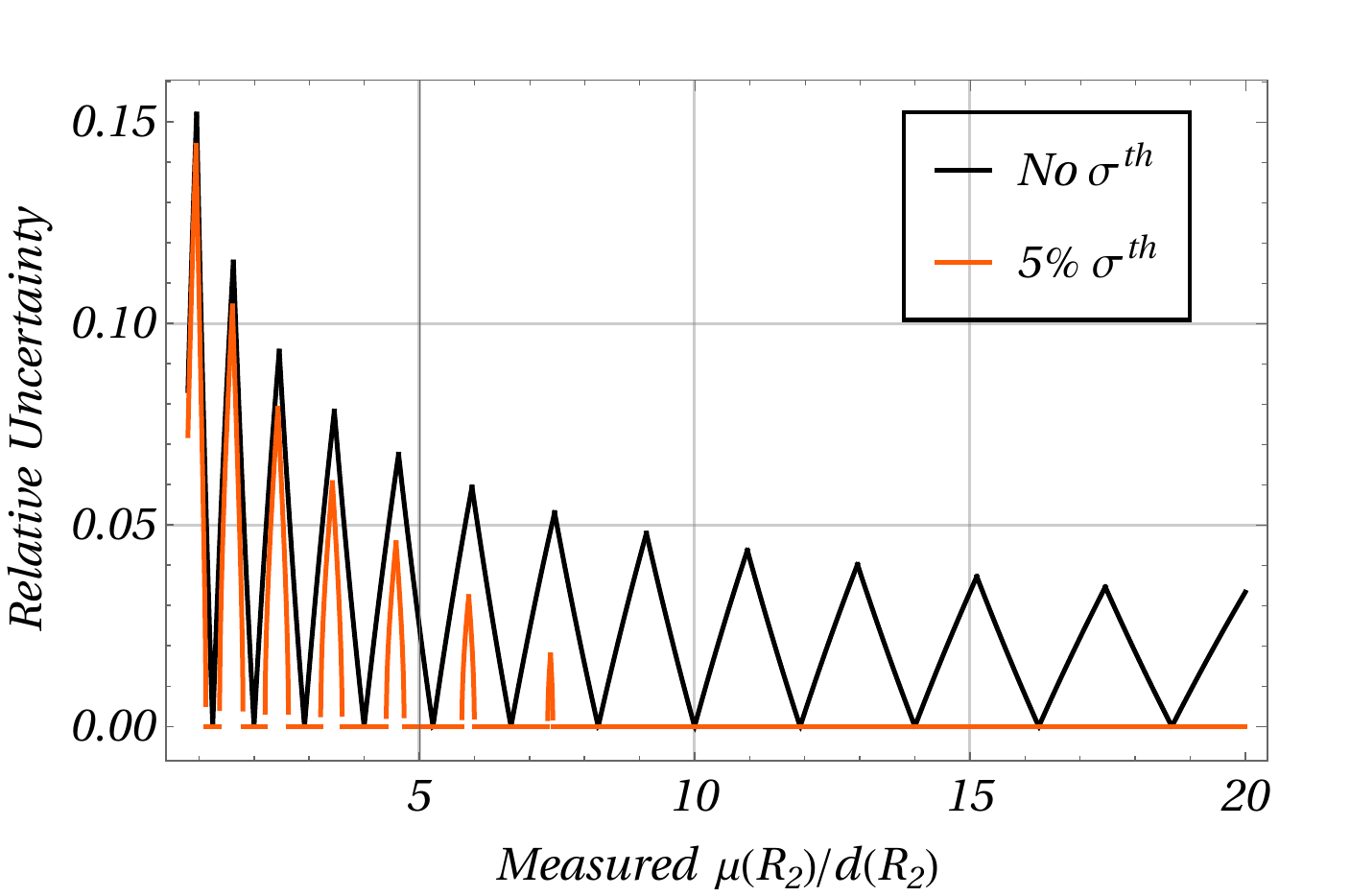}
    \caption{Left column: the ratios of the Dynkin index and dimension of the representation as a function of the dimension of the representation. The light red region in the upper left plot is the 2$\sigma$ band for the hypothetical future measurement centered at 0.25 with 10\% relative uncertainty.  
    Right column: the relative uncertainty needed to exclude the class of UV models realized by one scalar $\mathbb{Z}_6$ exotic particle for certain measured values of $\mu(R_3)/d(R_3)$ and $\mu(R_2)/d(R_2)$. For the black curves, we assume a vanishing theoretical uncertainty for the formulae in table.~\ref{tab:Ssolutions}. For the orange curves, we assume a symmetric 5\% relative theoretical uncertainty, denoted as $\sigma^{th}$.}
    \label{fig:muR3dR3_Y0}
\end{figure}

In the second step, one can check whether the model is excluded using the solutions for the quantum numbers of $R_2$ and $R_3$ obtained in the previous section. 
Due to the quantized nature of $R_2$ and $R_3$, the value of $\mu(R_2)/d(R_2)$,  $\mu(R_3)/d(R_3)$ and $F(R_2)$ are also quantized. 
Let us discuss the scalar model first as an example. 
We show the values of the ratios of the Dynkin indices and the dimensions for the $SU(3)_c$ representations in the upper left plot in figure.~\ref{fig:muR3dR3_Y0}, and the same for $SU(2)_L$ representations in the lower left plot in figure.~\ref{fig:muR3dR3_Y0}.
For both plots, we include the representations by requiring the Landau poles to be larger than 100 TeV, where we use the one-loop RG equations, and we assume that the mass of the complex scalar is at 1 TeV and it is either charged under $SU(3)_c$ or $SU(2)_L$.
Notice that one may push the Landau poles beyond 100 TeV with additional states in the UV. 
In general, the scale of the Landau pole depends on both the $SU(3)_c$ and $SU(2)_L$ representations, and we refer readers to the table.~9 in~\cite{DiLuzio:2015oha} for a qualitative two-loop running estimation. \footnote{Note that the masses of multiplets are assumed to be at $m_Z$ scale in~\cite{DiLuzio:2015oha}. Also see Refs.~\cite{Antipin:2017ebo, Antipin:2018zdg, Cacciapaglia:2019dsq} for some arguments on possible UV safety in the presence of large representations.}

Once a set of Wilson coefficients is measured in future experiments, one can calculate the combinations of the Wilson coefficients appearing on the right-hand side of the solutions (see e.g. in table~\ref{tab:Ssolutions}), which we call the ``measured value'' for the quantum numbers of $\phi$. Suppose the measured value of $\mu(R_3)/d(R_3)$ is 0.25 with 10\% uncertainty, of which the $2\sigma$ region is denoted by the light red band in the upper left plot, one can immediately falsify at 95\% CL the class of UV models realized by a $\mathbb{Z}_6$ exotic scalar. This is because no dot resides inside this band.
A similar argument can be made for $SU(2)_L$ representations. In practice, one should combine the information from the two plots to determine whether the $\mathbbm{Z}_6$ exotic models are excluded.
In the right column of figure.~\ref{fig:muR3dR3_Y0}, we also plot the relative uncertainties needed for exclusion with a specific measured value of $\mu(R_3)/d(R_3)$ and $\mu(R_2)/d(R_2)$. 
For example, dips in the black curve in the upper right plot correspond to the values of $\mu(R_3)/d(R_3)$ in the upper left plot, meaning that when the measured $\mu(R_3)/d(R_3)$ is close to one of the theoretical value, we can hardly exclude the class of UV models realized by one $\mathbb{Z}_6$ exotic scalar.

\begin{figure}[t]
    \centering
    \includegraphics[width = 7cm]{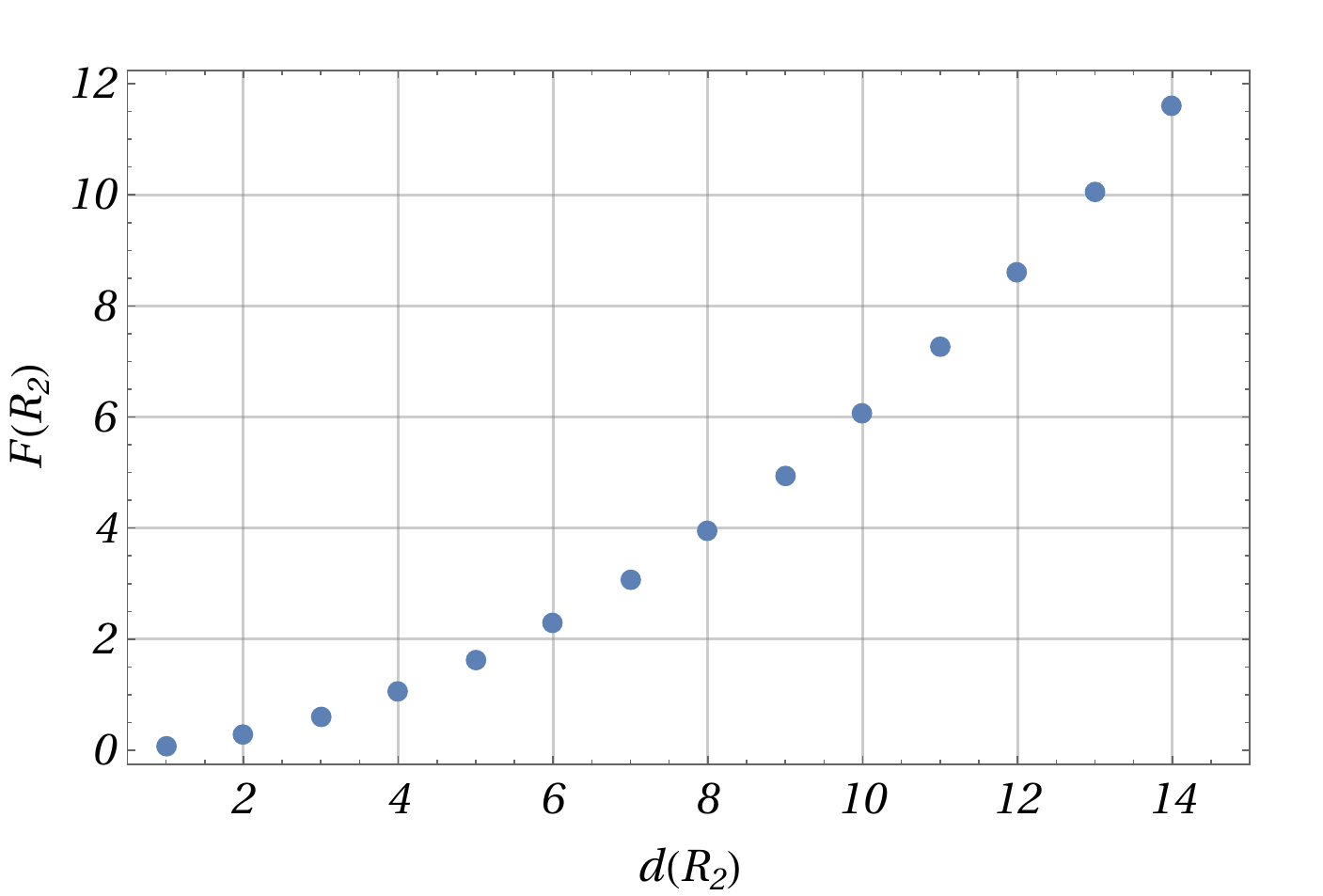}
    \includegraphics[width = 7cm]{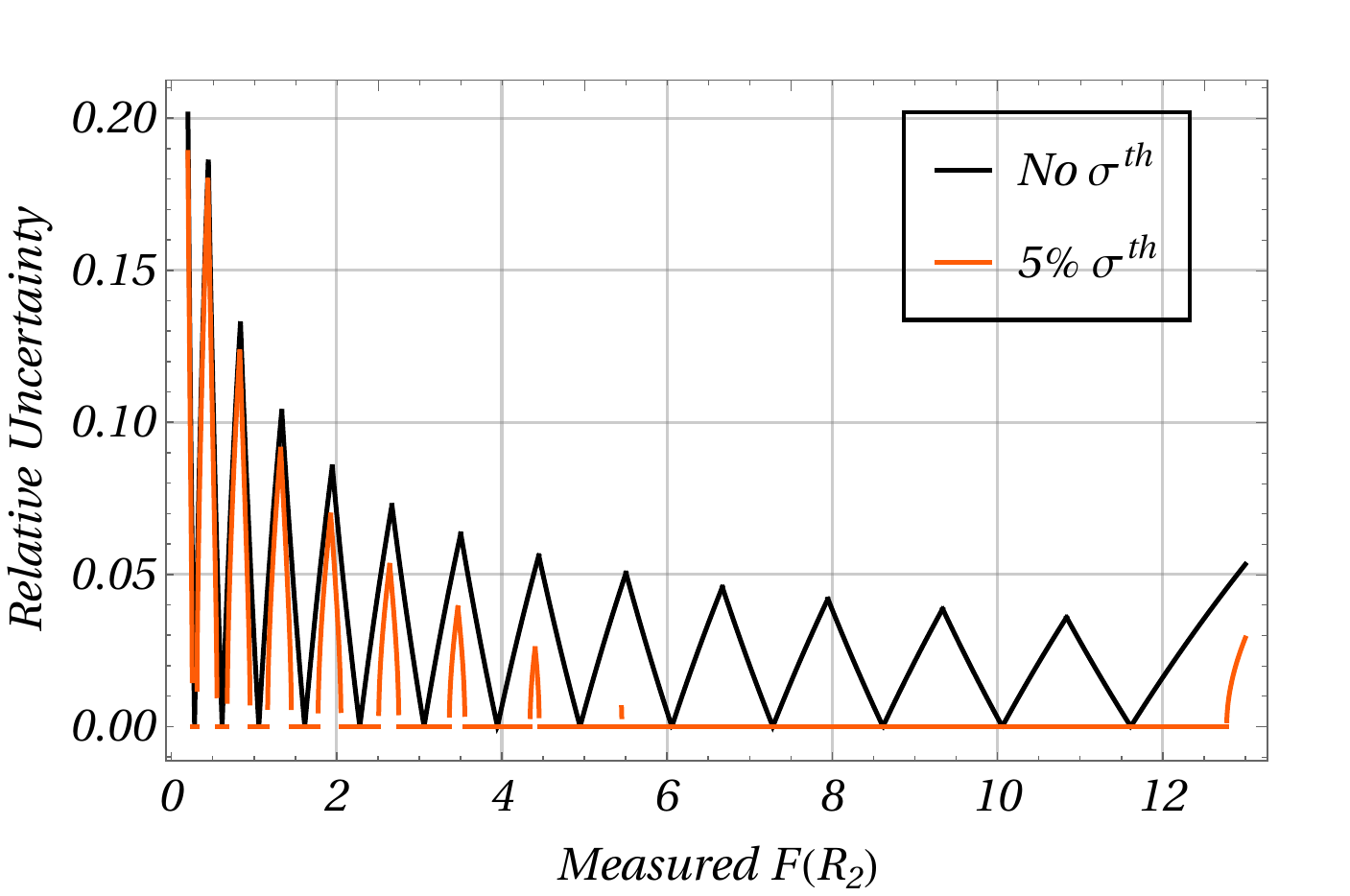}
     \includegraphics[width = 7cm]{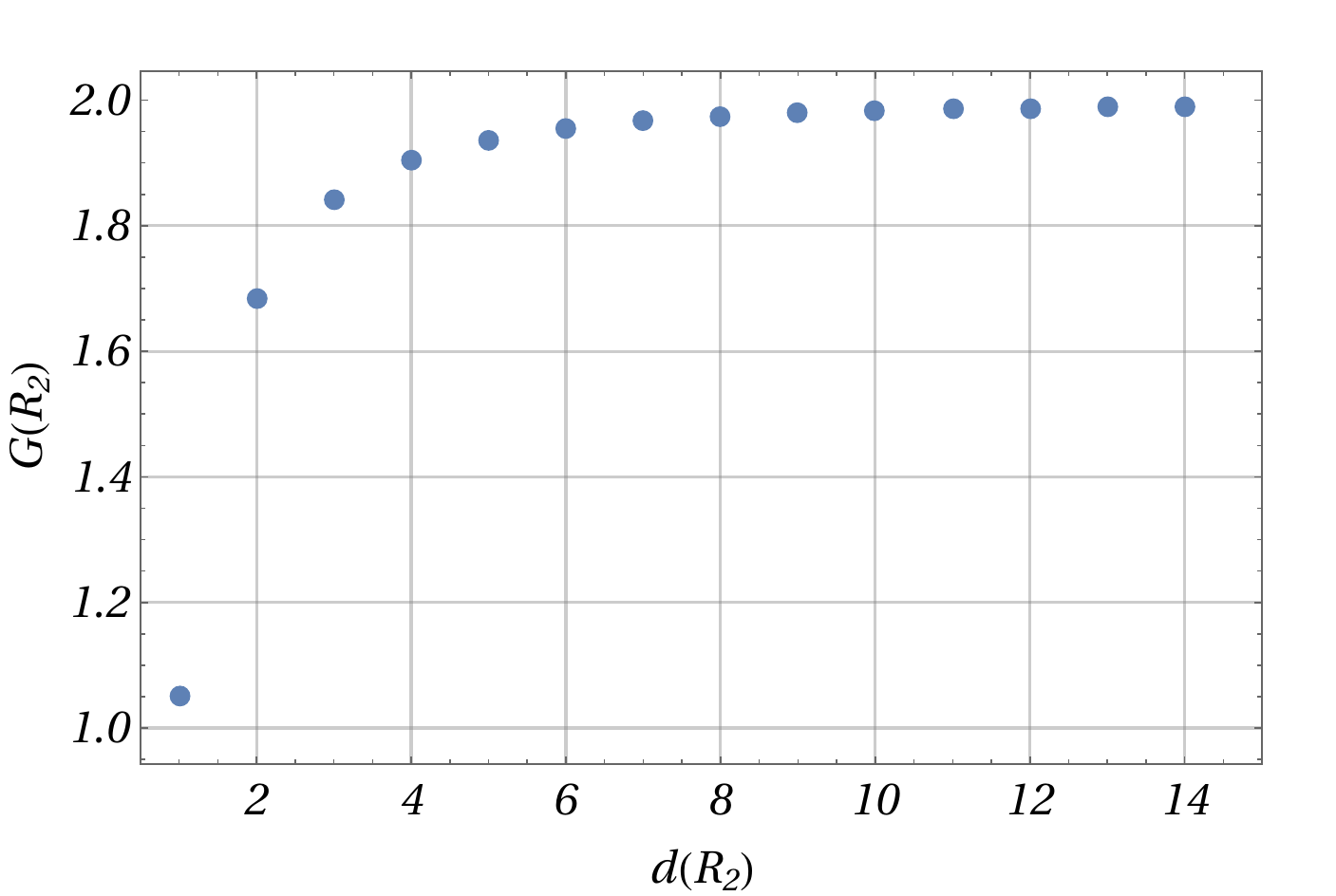}
    \includegraphics[width = 7cm]{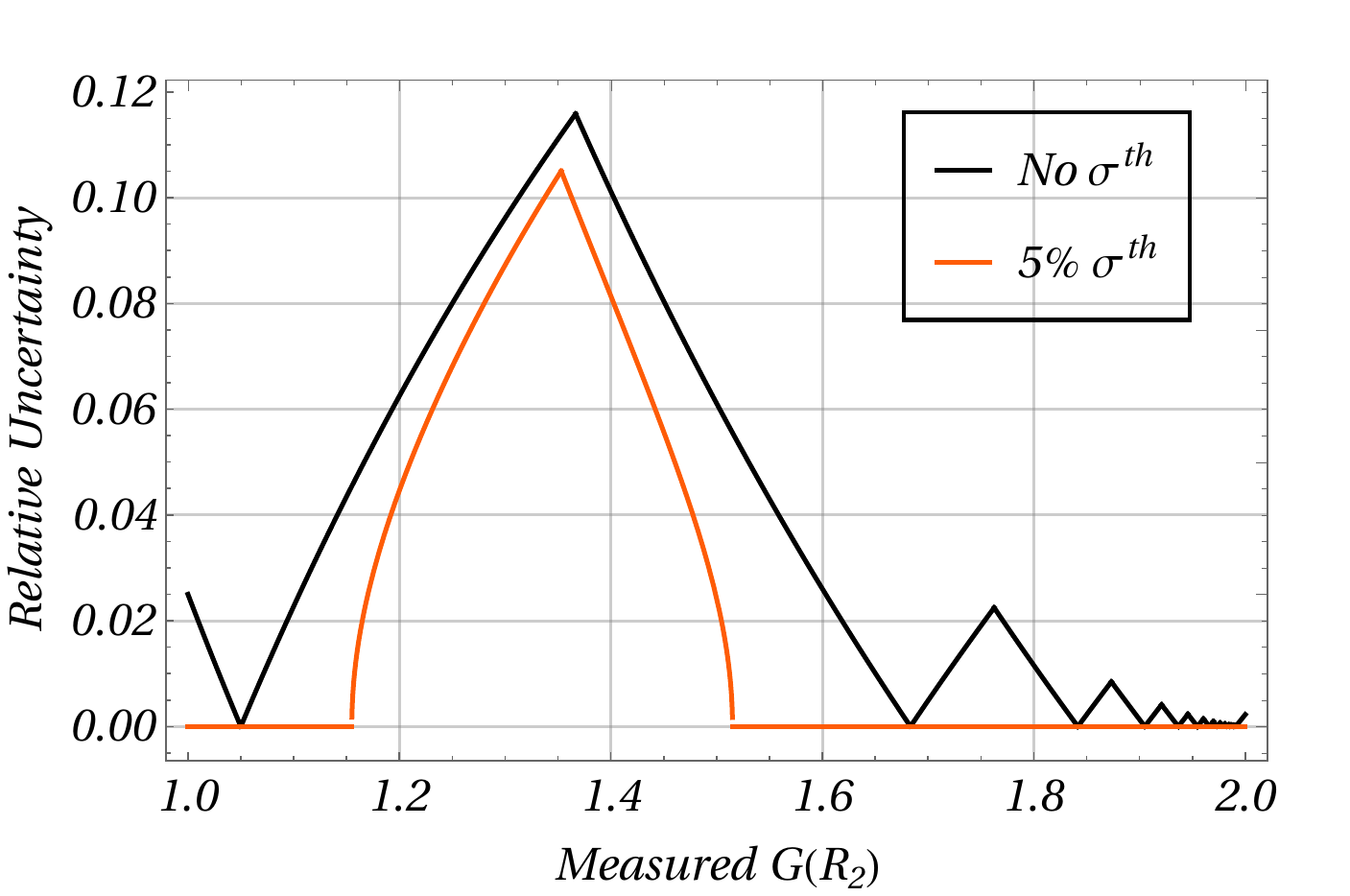}
    \caption{Left column: the group theoretical constants $F(R_2)$ and $G(R_2)$ as a function of the dimension of the representation $R_2$ of $\psi_1$. Right column: the relative uncertainty needed to make an exclusion for a specific measured value of $F(R_2)$ and $G(R_2)$ for different relative theoretical uncertainties denoted as $\sigma^{th}$.}
    \label{fig:F2}
\end{figure}

Similar arguments can be made for the fermion model. In this case, we do not have a simple solution to $\mu(R_2)/d(R_2)$, instead we have solutions to $F(R_2)$ and $G(R_2)$. In figure.~\ref{fig:F2}, we plot the possible values of $F(R_2)$ and $G(R_2)$ versus $d(R_2)$. 
Again if the measured value of $F(R_2)$ and $G(R_2)$ does not include any of the points in the plots within uncertainty, then one can falsify the the benchmark fermionic $\mathbb{Z}_6$ exotic model. 
More remarkably, one can see that the distribution of $F(R_2)$ is similar to that of $\mu(R_2)/d(R_2)$, while the value of $G(R_2)$ reaches a plateau as $d(R_2)$ increase. Therefore, any measured value of $G(R_2) > 2$ will immediately exclude the fermionic model. 

Here, we comment on the possible theoretical corrections to the formulae for the solutions of the quantum numbers. There are in general two types of corrections: the RG running of the Wilson coefficients and the higher loop corrections from matching. 
We argue that both of these corrections can be small, because, for $\mathbbm{Z}_6$ exotic UV theories, EFT operators are all generated at least at one-loop level, while the running effect is generally important when some tree-level generated Wilson coefficients contribute to the running of the loop generated ones. 
For the two-loop contribution, they should be relatively small as long as the UV model remains perturbative. We leave a quantitative estimation for future work.
Nevertheless, we demonstrate the impact of theoretical uncertainties (assumed to be 5\%) on the required experimental uncertainties for excluding the models; see the orange curves in the right plots in figure.~\ref{fig:muR3dR3_Y0} and \ref{fig:F2}. 
As one can see, the orange curves are lower than the black ones, meaning smaller experimental uncertainties are needed in general for exclusion, and there are also flat intervals centered around the theoretically allowed values, indicating the inability to make an exclusion when the measured center value is within the theoretical uncertainties.

In the third step, if one finds that the measured values of the aforementioned functions of $R_3$ and $R_2$ (such as $\mu(R_3)/d(R_3)$, etc) indeed cover some of their theoretical predictions in figure.~\ref{fig:muR3dR3_Y0} and figure.~\ref{fig:F2}, then one can use more correlations among the Wilson coefficients to verify whether the $\mathbb{Z}_6$ exotic models are indeed realized in nature.
In particular, one may find different solutions to the same quantum numbers involving different sets of Wilson coefficients. 
For example in the scalar model, $\mu(R_3)/d(R_3)$ can be solved as follows with only the Wilson coefficients of pure bosonic operators (but without using those of the four-fermion operators) if $\phi$ transform non-trivially under all the three gauge group factors:
\begin{eqnarray}
    \frac{\mu(R_3)}{d(R_3)} = -\frac{5 g_2 c_{3G} c_{HWB}^2 2 \lambda_{\mathbf{1}}^2}{2 c_{3W}c_{HB}g_3^3(c_{HB}g_1^2+5c_{HD}\lambda_{\mathbf{1}})}
    \text{ with }\lambda_{\mathbf{1}} = \frac{g_3 c_{HG}}{15 c_{3G}}=\frac{g_2 c_{HW}}{15 c_{3W}}\ .
    %-\frac{c^2_{HWB}c_{HG}^2}{5g_2^2g_3c_{3W}c_{HB}(6g_3 c_{HD}c_{HG}+g_1^2 c_{HB}c_{3G})}.
\end{eqnarray}
If the results obtained from different formulae match, then it will be strong evidence for the UV realized by the $\mathbbm{Z}_6$ exotic with Lagrangian in eq.~\eqref{eq:Lscalar}. 
For the fermionic model, the consistency between the values of the measured $F(R_2)$ and $G(R_2)$ can also be a sign of the UV model realized by the $\mathbbm{Z}_6$ exotic fermions as in eq.~\eqref{eq:Lfermion}.
A systematic way to find all possible solutions to the same quantum numbers will be explored in future work.

In the last step, after fixing the possible values of $R_2$ and $R_3$, one should use the formulae in the previous section to determine the ``measured values'' of the hypercharges $Y_{\phi,\psi}$. After that, one can use the criteria in eq.~(\ref{eq:quantization_Z6_main_1}) and (\ref{eq:quantization_Z6_main_2}) to determine whether the heavy particles transform as $\mathbb{Z}_6$ exotics.

%%%%%%%%%%%%%%%%%%%%%%%%%%%%%%%%%%%%%%%%%%%%%%%%%%%%%%%%%%%%
\section{Non-decoupling scalars}
\label{sec:non-decoupling_case}
%%%%%%%%%%%%%%%%%%%%%%%%%%%%%%%%%%%%%%%%%%%%%%%%%%%%%%%%%%%%

There is another class of minimal BSM extensions where the new particles cannot decouple from the weak scale. This happens when there are additional sources of EWSB other than the SM Higgs doublet. In this section, we aim to understand whether the scalars triggering EWSB can be $\mathbb{Z}_6$ exotic.
Notice that SMEFT cannot be used when the new particles do not decouple from the weak scale. We refer the readers to e.g.~\cite{Cohen:2020xca, Banta:2021dek} for more discussions on the non-decoupling nature of the scalars triggering EWSB. 

To avoid breaking $SU(3)_c$ at the weak scale, these scalars are singlets under $SU(3)_c$, which implies that $\mathcal{N}(R_3)=0$ in eq.~(\ref{eq:quantization_Z6_main_1}). Nevertheless, they can be in nontrivial representations $(R_2, Q_Y)$ under $SU(2)_L\times U(1)_Y$. It is convenient to label $R_2$ using the spin $j$ of the corresponding representation under $SU(2)_L$ and the N-ality $\mathcal{N}(R_2)=2j$, where $j$ is an integer or half-integer.

Let us consider the most general EWSB sector without assuming custodial symmetry. The scalars that we are interested in are classified in representations $(j, Q_Y)$ under $SU(2)_L\times U(1)_Y$. 
In the gauge eigenstate, one has the general kinetic terms 
\bea
\mathcal{L}_{\text{kin}}=\sum_{i} c_i | D_i^\mu \phi_i |^2\ , 
\eea
where $c_i=1$ for complex scalars, $c_i=\frac{1}{2}$ for real scalars, and the index $i$ runs over all the flavors of scalars. 
To trigger EWSB (but without breaking $U(1)_{\text{EM}}$), there must exist an electric neutral component in each $\phi_i$ with a non-vanishing vacuum expectation value, this implies that the quantum numbers $(j, Q_Y)$ are subject to the constraints as
\bea
-j\leq Q_Y \leq j \quad\quad\text{and}\quad\quad j+Q_Y\in \mathbb{Z} \ .
\label{eq:electric_neutral_condition}
\eea
The reason is that $j+Q_Y$ gives the largest possible electric charge for a component in the EW multiplet, which needs to be a positive integer. Moreover, $-j+Q_Y$ gives the lowest possible electric charge, which is automatically integer-valued if $j+Q_Y$ is already an integer, and it has to be negative to accommodate an electric neutral component.  
From eq.~(\ref{eq:electric_neutral_condition}), we see that $Q_Y$ has to be an integer when $j$ is an integer, and $Q_Y$ is a half-integer when $j$ is a half-integer. Consequently, eq.~(\ref{eq:quantization_Z6_main_1}) is automatically satisfied since 
\bea
0=6 Q_Y\ \text{mod}\ 3\ , 
\eea
where we have used the fact that $Q_Y$ is either an integer or a half-integer. Furthermore, eq.~(\ref{eq:quantization_Z6_main_2}) is also satisfied because
\bea
2j-6Q_Y= 2 (j-Q_Y) -4 Q_Y=0 \ \text{mod}\ 2\ ,
\eea
where we have used the fact that $j-Q_Y$ is an integer and $0=4 Q_Y\ \text{mod}\ 2$. Hence these scalars cannot be $\mathbb{Z}_6$ exotics.

To summarize, the scalars triggering EWSB must be invariant under the $\mathbb{Z}_6$ subgroup of $\tilde{G}=SU(3)_c\times SU(2)_L\times U(1)_Y$, i.e. these particles cannot be $\mathbb{Z}_6$ exotics. This is because the scalar multiplet must contain an electric neutral component. (For the same reason, the EW multiplets in all the minimal dark matter models are $\mathbb{Z}_6$ invariant. See e.g.~\cite{Bottaro:2022one} and the references therein for concrete examples.) Finally, we close this section by giving some concrete examples of the scalars in the EWSB sector. In our notation of $(j, Q_Y)$, the SM Higgs doublet is $(\frac{1}{2}, \frac{1}{2})$, the real and complex triplets in Georgi-Machacek Model~\cite{Georgi:1985nv} are $(1, 0)$ and $(1, 1)$, and a Higgs septet~\cite{Harris:2017ecz} has quantum numbers $(3, 2)$. All these scalars are invariant under the $\mathbb{Z}_6$ group, c.f. eqs.~(\ref{eq:quantization_Z6_main_1}) and~(\ref{eq:quantization_Z6_main_2}).

%%%%%%%%%%%%%%%%%%%%%%%%%%%%%%%%%%%%%%%%%%
\section{Discussions and outlook}
\label{sec:conclusion}
%%%%%%%%%%%%%%%%%%%%%%%%%%%%%%%%%%%%%%%%%%%
As particle phenomenologists motivated by generalized symmetries, we reconsider the precise form of the SM gauge group, which has important implications for the allowed representations of the heavy particles in the UV. 
At the intuitive level, there are natural connections between heavy particles, Wilson line operators with one-form global symmetries acting on them, and high dimensional operators induced at IR. Since the Wilson lines can be screened by creating particle-antiparticle pairs from the vacuum, the line operators are only well-defined below the mass threshold of the heavy particles. Saying it differently, there is electric one-form global symmetry acting on the Wilson lines at low energy, but the electric one-form symmetry is explicitly broken above the mass threshold of heavy particles. Hence, measuring the mass scale at which the lightest $\mathbb{Z}_6$ exotic particle appears is just measuring the scale at which the corresponding electric one-form symmetry gets explicitly broken.

To determine what is the SM gauge group, we study the heavy particles not invariant under the $\mathbb{Z}_6$ subgroup of $SU(3)_c\times SU(2)_L\times U(1)_Y$, i.e. the $\mathbb{Z}_6$ exotics, from the perspective of SMEFT. 
Our main results can be summarized as follows.
\begin{itemize}
    \item We demonstrate that the $\mathbb{Z}_6$ exotics cannot appear in tree-level UV completions of SMEFT in weakly coupled theories. This result is consistent with all the examples in~\cite{Li:2022abx, Li:2023cwy, Li:2023pfw}. 

   \item At the one-loop level, we demonstrate a strategy to examine the UV models involving $\mathbb{Z}_6$ exotics. The idea is to extract the quantum numbers of heavy particles from the Wilson coefficients, and the formulae of which can be obtained if one leaves the quantum numbers of the UV particle as undetermined variables during matching.
   Our analysis gives stronger motivation for studying one-loop matching in SMEFT, since it is mandatory to identify the $\mathbb{Z}_6$ exotic particles and hence to determine the SM gauge group. A systematic study of the loop-level dictionary between the Wilson coefficients of the high dimensional operators and UV models involving $\mathbb{Z}_6$ exotic particles is warranted. 
   
   \item When the heavy particles do not satisfy the decoupling limit, SMEFT cannot be used. In this class of non-decoupling models, we prove that all the scalars that can trigger electroweak symmetry breaking cannot be $\mathbb{Z}_6$ exotic. This justifies the quantum number of the SM Higgs doublet. 
\end{itemize}

There are a few future directions. 
\begin{itemize}
\item The spectrum of the allowed representations for $\mathbb{Z}_6$ exotic particles is extremely broad. If these particles have a decoupling limit and they are much higher than the weak scale, SMEFT serves as a universal tool to probe these particles. However, when the mass is small, the search strategy can vary depending on the specific quantum numbers of $\mathbb{Z}_6$ exotic particles. A better understanding is needed for determining the optimal model-independent strategy for searching the light $\mathbb{Z}_6$ exotic particles. 

In the presence of very light $\mathbb{Z}_6$ exotic particles, the corresponding Wilson lines are screened. Saying it differently, the corresponding electric one-form symmetry is explicitly broken above the mass threshold of light $\mathbb{Z}_6$ exotic particles.

\item We notice that the lightest $\mathbb{Z}_6$ exotic particle is a cosmologically stable relic. As we discussed in section.~\ref{sec:non-decoupling_case}, if it is a $SU(3)_c$ singlet then it must contain electric charge and thus is strongly constrained by the direct terrestrial search experiments~\cite{Perl:2009zz} and multiple cosmological observations (e.g. CMB anisotropy and matter power spectrum) due to their interaction with ordinary baryons in the early universe~\cite{Kovetz:2018zan, Dubovsky:2003yn, dePutter:2018xte, Xu:2018efh, Buen-Abad:2021mvc}. 
On the other hand, if the lightest $\mathbb{Z}_6$ exotic is charged under $SU(3)_c$, then it will form into hadrons after the QCD phase transition. Such exotic hadrons also receive very tight constraints from various astrophysical considerations~\cite{Dimopoulos:1989hk, Chuzhoy:2008zy, Hertzberg:2016jie, Gould:1989gw, Mack:2007xj}.
Therefore, to avoid overproduction in the early universe, one expects the reheating temperature to be much smaller than the mass of those $\mathbb{Z}_6$ exotics, unless there is a mechanism to enhance their annihilation rates.
This implies a lower bound on the masses of the $\mathbb{Z}_6$ exotics, because the reheating temperature cannot be lower than the characteristic temperature of the Big Bang Nucleosynthesis. See e.g.~\cite{Gan:2023jbs} for a related study on milli-charged particles.
While most of the aforementioned cosmological and astrophysical bounds on the relic abundance of electrically charged particles apply to a mass range from sub-MeV to ${\cal O}(100)$ GeV,  bounds for $\mathbb{Z}_6$ exotics of multi-TeV scale remain to be explored. Furthermore, the studies on the interactions of the hadrons formed by colored massive stable particles are still preliminary~\cite{Dover:1979sn, Nardi:1990ku, Arvanitaki:2005fa, Kang:2006yd, Jacoby:2007nw}, and a more quantitative study will help to determine the bounds on the reheating temperature, the masses, and the representations of the $\mathbb{Z}_6$ exotic particles. 

\item It is certainly warranted to investigate the phenomenology of $\mathbb{Z}_6$ exotic particles at future colliders. The current constraints on the mass of stable color singlet charged particles are around $1\sim 2$ TeV for integer charged particles~\cite{ATLAS:2023zxo}, and around $600$ GeV for fractionally charged particles~\cite{CMS-PAS-EXO-19-006}, and both of them rely on the characteristic ionization energy loss $dE/dx$ in the detector to discriminate signal events from backgrounds.
On the other hand, for colored $\mathbb{Z}_6$ exotic, there are large uncertainties in the modeling of parton shower and hadronization processes, and the modeling of exotic hadron interaction with detectors as well. 
In the future, one might be interested in analyzing the prospective sensitivity for probing $\mathbb{Z}_6$ exotics in various proposed collider experiments, such as the muon collider.
\end{itemize}

As shown in literature, generalized symmetry is a powerful tool that leads us to develop a more coherent, unified, and deeper understanding of various known physics across many frontiers. This work is another piece of this kind. However, we are optimistic that eventually generalized symmetries can offer striking new solutions to the outstanding problems in particle physics.   

%%%%%%%%%%%%%%%%%%%%%%%%%%%%%%%%%%%%%%%
\vspace{0.1in}
\noindent
{\bf Acknowledgements.---}
%%%%%%%%%%%%%%%%%%%%%%%%%%%%%%%%%%%%%%%%%%%%
We thank Marco Costa and Andrea Luzio for illuminating discussions, and Bobby Acharya, C\'eline Degrande, Gauthier Durieux, Mehrdad Mirbabayi, Rudin Petrossian-Byrne, Sharam Vatani, Luca Vecchi, Giovanni Villadoro for their helpful comments and feedback. We also thank one anonymous referee for many constructive suggestions.
H.-L.L. is supported by the postdoctoral fellowship FSR of Universite Catholique de Louvain.  
The work of L.X.X. is partially supported by ERC grant n.101039756.

\vspace{0.1in}
\noindent {\bf Note Added.---}
When this work is finalized, a similar work~\cite{Alonso:2024pmq} appears on arXiv. Although the motivation is similar, we focus more on the implications of the heavy $\mathbb{Z}_6$ exotics in SMEFT and emphasize the importance of loop-level matching to probe the $\mathbb{Z}_6$ exotic particles. 
After our paper was posted on arXiv, another paper~\cite{Koren:2024xof} appeared, which discusses fractional-charged particles from a complementary perspective. 

\appendix
%%%%%%%%%%%%%%%%%%%%%%%%%%%%%%%%%%%%%%%%%%%%%%%%%%%%%
\section{Review on the $\mathbb{Z}_6$ group and how it acts on the particles}
\label{app1:Z6}
%%%%%%%%%%%%%%%%%%%%%%%%%%%%%%%%%%%%%%%%%%%%%%%%%%%%

%%%%%%%%%%%%%%%%%%%%%%%%%%%%%%%%%%%%%%%%%%%%
\subsection{$\Gamma=\mathbb{Z}_6$}
%%%%%%%%%%%%%%%%%%%%%%%%%%%%%%%%%%%%%%%%%%

In this appendix, we review how the $\mathbb{Z}_6$ subgroup of $\tilde{G}=SU(3)_c\times SU(2)_L \times U(1)_Y$ acts on the field in representation $(R_3, R_2, Q_Y)$ of $\tilde{G}$. 

The $\mathbb{Z}_6$ subgroup has 6 group elements $\{\alpha, \alpha^2, \alpha^3, \alpha^4, \alpha^5, \alpha^6=1\}$, where 
\bea
\alpha=\left( e^{\frac{2 \pi i}{3}} \mathbbm{1}_{3\times 3}, \ e^{\pi i} \mathbbm{1}_{2\times 2}, \ e^{\frac{2 \pi i}{6}} \right)\ ,
\eea
and it acts on a field in the representation $(R_3, R_2, Q_Y)$ as
\bea
U_\alpha(R_3,R_2,Q_Y)=e^{\frac{2 \pi i}{3} \mathcal{N}(R_3)+i \pi \mathcal{N}(R_2)+\frac{2\pi i}{6} (6 Q_Y)}=e^{2\pi i \left(\frac{\mathcal{N}(R_3)}{3}+\frac{\mathcal{N}(R_2)}{2}+Q_Y\right)}\ ,
\eea
where $\mathcal{N}(R_3)$ and $\mathcal{N}(R_2)$ are the numbers of boxes of the Young diagrams for the $SU(3)_c$ representation $R_3$ and the $SU(2)_L$ representation $R_2$, respectively. The generator of $U(1)_Y$ is taken to be $6 Q_Y$ which has to be integer-valued. Notice that this choice for the normalization of the hypercharge is motivated by the fact that the left-handed quark doublet has charge $+1/6$.

The group element $\alpha$ acts trivially on $(R_3, R_2, Q_Y)$ when $U_\alpha(R_3,R_2,Q_Y)=1$, i.e. 
\bea
\frac{\mathcal{N}(R_3)}{3}+\frac{\mathcal{N}(R_2)}{2}+Q_Y \in \mathbb{Z}. 
\label{eq:quantization_Z6_1}
\eea
One can rewrite $\mathcal{N}(R_3)= 6 Q_Y+x$ and $\mathcal{N}(R_2)= 6 Q_Y+y$, where $x$ and $y$ are also integer-valued. Then eq.~(\ref{eq:quantization_Z6_1}) becomes
\bea
6 Q_Y+\frac{x}{3}+\frac{y}{2}\in \mathbb{Z}\ ,
\eea
which implies $x=0 \ \text{mod}\  3$ and $y=0 \ \text{mod}\  2$. Consequently, we obtain the quantization conditions for $\mathbb{Z}_6$ acting trivially on $(R_3, R_2, Q_Y)$:
\bea
\mathcal{N}(R_3)&=& 6 Q_Y\ \text{mod}\ 3\ ,\\
\mathcal{N}(R_2)&=& 6 Q_Y\ \text{mod}\ 2\ .
\label{eq:quantization_Z6_2}
\eea
It is easy to check that all the SM fields satisfy these conditions, hence $\mathbb{Z}_6$ acts trivially. 

%%%%%%%%%%%%%%%%%%%%%%%%%%%%%%%%%%%%%%%%%%%%%%%%%%%%
\subsection{$\Gamma=\mathbb{Z}_3$}
%%%%%%%%%%%%%%%%%%%%%%%%%%%%%%%%%%%%%%%%%%%%%%%%%%%%%

Similarly, one can work out the quantization condition for $\mathbb{Z}_3$ acting trivially on $(R_3, R_2, Q_Y)$. 

As a subgroup of $\mathbb{Z}_6$, $\mathbb{Z}_3$ has 3 group elements $\{\alpha^2, \alpha^4, \alpha^6=1\}$, where the group element $\alpha^2$ acts on $(R_3, R_2, Q_Y)$ as
\bea
U_{\alpha^2}(R_3,R_2,Q_Y)=e^{\frac{4 \pi i}{3} \mathcal{N}(R_3)+ 2\pi i \mathcal{N}(R_2)+\frac{2\pi i}{3} (6 Q_Y)}=e^{2\pi i \left(\frac{2 \mathcal{N}(R_3)}{3}+2 Q_Y\right)}\ .
\eea
Hence $\mathbb{Z}_3$ acts trivially on $(R_3, R_2, Q_Y)$ when $U_{\alpha^2}(R_3,R_2,Q_Y)=1$, i.e.
\bea
\frac{2 \mathcal{N}(R_3)}{3}+2 Q_Y \in \mathbb{Z}. 
\eea
Using the parametrization $\mathcal{N}(R_3)= 6 Q_Y+x$ with $x$ being integer-valued, one obtains $x=0 \ \text{mod}\  3$. Consequently, the quantization condition for $\mathbb{Z}_3$ acting trivially on $(R_3, R_2, Q_Y)$  is:
\bea
\mathcal{N}(R_3)= 6 Q_Y\ \text{mod}\ 3\ ,
\eea
while $R_2$ can take any irreducible representation of $SU(2)_L$.

%%%%%%%%%%%%%%%%%%%%%%%%%%%%%%%%%%%%%%%%%%%%%%
\subsection{$\Gamma=\mathbb{Z}_2$}
%%%%%%%%%%%%%%%%%%%%%%%%%%%%%%%%%%%%%%%%%%%%%%%%%
At last, we work out the quantization condition for $\mathbb{Z}_2$ to act trivially. 

As a subgroup of $\mathbb{Z}_6$, $\mathbb{Z}_2$ has 2 group elements $\{\alpha^3, \alpha^6=1\}$, where the group element $\alpha^3$ acts on $(R_3, R_2, Q_Y)$ as
\bea
U_{\alpha^3}(R_3,R_2,Q_Y)=e^{\frac{6 \pi i}{3} \mathcal{N}(R_3)+ 3\pi i \mathcal{N}(R_2)+\pi i (6 Q_Y)}=e^{2\pi i \left(\frac{\mathcal{N}(R_2)}{2}+3 Q_Y\right)}\ .
\eea
Hence $\mathbb{Z}_2$ acts trivially on $(R_3, R_2, Q_Y)$ when $U_{\alpha^3}(R_3,R_2,Q_Y)=1$, i.e.
\bea
\frac{\mathcal{N}(R_2)}{2}+3 Q_Y \in \mathbb{Z}. 
\eea
Using the parametrization $\mathcal{N}(R_2)= 6 Q_Y+y$ with $y$ being integer-valued, one obtains $y=0 \ \text{mod}\  2$. Consequently, the quantization condition for $\mathbb{Z}_2$ acting trivially on $(R_3, R_2, Q_Y)$  is:
\bea
\mathcal{N}(R_2)= 6 Q_Y\ \text{mod}\ 2\ ,
\eea
while $R_3$ can take any irreducible representation of $SU(3)_c$.

%%%%%%%%%%%%%%%%%%%%%%%%%%%%%%%%%%%%%%%%%%%%%%%%%%%
\section{Analytical expressions for Dynkin indices and dimensions}
\label{app:derivation}
%%%%%%%%%%%%%%%%%%%%%%%%%%%%%%%%%%%%%%%%%%%%%%%%%%%%

If we use the numbers of boxes in each row $[l_1, l_2]$ and $[l]$ of the Young diagrams to represent the irreducible representations of $SU(3)_c$ and $SU(2)_L$, we have the following formulae for the Dynkin indices $\mu(R_i)$ and the dimensions $d(R_i)$:
\begin{eqnarray}
    \mu([l_1,l_2])&=&\frac{1}{48} (2 + l_1) (1 + l_1 - l_2) (1 +l_2) \nn \\
    &\ & ( l_1^2 +l_1(3-l_2) + l_2^2),\\
    d([l_1,l_2])&=&\frac{1}{2} (2 + l_1) (1 + l_1 - l_2) (1 + l_2) ,\\
   \mu([l])&=&\frac{1}{12}l(l+1)(l+2),\\
   d([l])&=&l+1 .   
\end{eqnarray}

%%%%%%%%%%%%%%%%%%%%%%%%%%%%%%%%%%%%%%%%%%%%%%%%%%%%%%%%%%%%%%%%%%
\bibliographystyle{JHEP}
\bibliography{SM_gauge_group}
%%%%%%%%%%%%%%%%%%%%%%%%%%%%%%%%%%%%%%%%%%%%%%%%%%%%%%%%%%%%%%%%%

\end{document}